\documentclass[12pt]{article}
\usepackage{a4wide,epsfig,amsmath,amssymb,cite,scalefnt}
\def\be{\begin{equation}}
\def\ee{\end{equation}}
\def\besub{\begin{subequations}}
\def\eesub{\end{subequations}}
\def\bea{\begin{eqnarray}}
\def\eea{\end{eqnarray}}
\def\Tr{{\rm Tr }}

\def\frak#1#2{{\textstyle{\frac{#1}{#2}}}}

\def\Htil{\tilde H}

\def\lfp{8\pi^2}

\def\nn{\nonumber\\}
\def\aep{\alpha_e}
\def\as{\alpha_s}

\newcommand{\epscalar}{$\varepsilon$-scalar}
\newcommand{\epscalars}{$\varepsilon$-scalars}
\newcommand{\reference}[1]{Ref.~\cite{#1}}
\newcommand{\eqn}[1]{Eq.~(\ref{#1})}
\newcommand{\abbrev}{\scalefont{.9}}
\newcommand{\drbar}{$\overline{\mbox{\abbrev DR}}$}
\newcommand{\msbar}{$\overline{\mbox{\abbrev MS}}$}
\newcommand{\asDRbar}{\alpha_s}
\newcommand{\asMSbar}{\alpha_s^{\overline{\rm MS}}}
\newcommand{\betaDRbar}{\beta^{\overline{\rm DR}}}
\newcommand{\betaMSbar}{\beta^{\overline{\rm MS}}}
\newcommand{\gammaDRbar}{\gamma^{\overline{\rm DR}}}
\newcommand{\gammaMSbar}{\gamma^{\overline{\rm MS}}}
\newcommand{\ZsDRbar}{Z_s^{\overline{\rm DR}}}

\newcommand{\ZmDRbar}{Z_m^{\overline{\rm DR}}}

\newcommand{\qcd}{{\abbrev QCD}}
\newcommand{\sqcd}{{\abbrev SQCD}}

\newcommand{\mssm}{{\abbrev MSSM}}
\newcommand{\dreg}{{\abbrev DREG}}
\newcommand{\dred}{{\abbrev DRED}}
\newcommand{\nsvz}{{\abbrev NSVZ}}
\newcommand{\mDRbar}{m^{\overline{\rm DR}}}
\newcommand{\mMSbar}{m^{\overline{\rm MS}}}
\newcommand{\apiDR}{\frac{\alpha_s}{\pi}}
\newcommand{\apiMS}{\frac{\asMSbar}{\pi}}
\newcommand{\aepi}{\frac{\alpha_e}{\pi}}
\newcommand{\api}{\frac{\alpha_s}{\pi}}

\def \qq{\qquad}  
\def\pa{\partial}
\def\ep{\epsilon}
\def \la{\lambda}

\def\Hbar{{\overline{H}}}

\def\alphadot{\dot\alpha}

\def\sy{supersymmetry}
\def\sic{supersymmetric} 
\def\psib{\overline{\psi}}
\def\alphab{\overline{\alpha}}
\def\betab{\overline{\beta}}
\def\deltab{\overline{\delta}}
\def\gammab{\overline{\gamma}}
\def\sigb{\overline{\sigma}}
\def\sigp{\sigma^{\prime}}
\def\semi{;\hfil\break}
\begin{document}

\title{\vskip-3cm{\baselineskip14pt
    \begin{flushleft}
      \normalsize LTH751  \\
 \normalsize TTP/07-17\\
 \normalsize  SFB/CPP-07-40\\
 \normalsize  NSF-KITP-07-153\\
\end{flushleft}}
  \vskip2cm
The four-loop \dred\ gauge $\beta$-function and 
fermion mass anomalous dimension for general gauge groups
}
\author{\small Ian Jack$^{(a)}$,
  D.R. Timothy Jones$^{(a)}$,
  Philipp Kant$^{(b)}$,
  Luminita Mihaila$^{(b)}$\\[1em]
  {\small\it (a) Department of Mathematical Sciences,
    University of Liverpool,}\\
  {\small\it Liverpool L69 3BX, UK}\\
  {\small\it (b) Institut f{\"u}r Theoretische Teilchenphysik,
    Universit{\"a}t Karlsruhe,}\\
{\small\it 76128 Karlsruhe, Germany}\\
}  
\date{}
  
\maketitle
 
\thispagestyle{empty}

\begin{abstract}
We present four-loop results for the gauge $\beta$-function and 
the fermion mass anomalous dimension for a gauge theory 
with a general gauge group and a multiplet of fermions 
transforming according to an arbitrary representation, calculated 
using the dimensional reduction scheme. In the 
special case of a supersymmetric theory we confirm 
previous calculations of both the gauge $\beta$-function and 
the gaugino mass $\beta$-function. 


\end{abstract}

\vfill


\section{Introduction}

In recent papers some of us  presented calculations of the \qcd{}
$\beta$-function, $\beta_s$,  and the fermion mass anomalous dimension
(or  mass $\beta$-function), $\gamma_m$  through three
loops~\cite{Harlander:2006rj}  and four loops~\cite{Harlander:2006xq}
using  the \dred{} (or \drbar)  scheme, which is based on 
regularisation by dimensional reduction~\cite{siegel,Capper}. 
An interesting  feature of these calculations is the
dependence of $\beta_s$ on the evanescent  couplings: \epscalar{}
interactions that do not renormalise like  the gauge  coupling. At three
loops $\beta_s$ depends on the \epscalar{} Yukawa coupling,  and at four
loops it also depends on the \epscalar{} quartic interaction.  The
first explicit calculations of the one loop corrections to this quartic
interaction appeared in  \reference{Jack:1993ws} (for a particular
$SU(2)$ model), and in \reference{Harlander:2006xq} (for \qcd). Here
we generalise  the calculation to $SU(N)$, $SO(N)$ and $Sp(N)$. This
involves some quite  interesting and (relatively) little known group
theory. We also  similarly generalise the result from
\reference{Harlander:2006xq} for the \epscalar{} Yukawa coupling.

Of course it is the $SU(3)$ case described in the previous papers which
is  most obviously currently useful, but the general result  is also of
interest, for possible future applications to other symmetry  groups,
and if only as  a further  test of the validity of the \dred\ procedure.
Our confidence in  this is reinforced by once again comparing our
results for the special case of  supersymmetry (when there is a single
fermion  multiplet in the adjoint representation). The result for
$\beta_s$ for a renormalisable  ${\cal N} =1$ \sic\ theory was given
through four loops in Ref.~\cite{Jack:1998uj}, the derivation  being
based on  the completion of a construction of the coupling constant
redefinition  connecting the  \dred\ scheme to the \nsvz\ scheme
developed in Ref.~\cite{cjn}. Here we not only verify this result
through  four loops (in the special case of a theory with no
superpotential) but also we verify the  result for the gaugino
$\beta$-function through the same order.  This is of interest because of
course  the gaugino mass breaks supersymmetry, and the issue of
regularisation and renormalisation  of softly-broken \sic\ theories
present additional subtleties.   The exact formula for the gaugino
$\beta$-function (expressing it in terms of  $\beta_s$), as first
derived in Ref.~\cite{Jack:1997pa}\ (inspired by an observation by
Hisano and Shifman in Ref.~\cite{HS}) relied heavily on the spurion 
formalism, as developed in particular by Yamada~\cite{yam}; it is
reassuring to find that the relationship between the two
$\beta$-functions  indeed holds in an explicit \dred\ calculation.

In section~\ref{sec::rev}\ we review the renormalisation procedure  for
a gauge theory using \dred; then in section~\ref{sec::epsfour}\  we
describe the one loop renormalisation of the \epscalar{}
self-interaction.  We first give results for the $SU(N)$ case,
explaining how to reduce  to the special cases $N=2$ and $N=3$.  We 
then generalise to expressions valid for an arbitrary groups.

In section~\ref{sec::betafour}\ we give the full four-loop  results
for $\beta_s$ and $\gamma_m$ for the general case, and  in
section~\ref{sec::moresusy}\ we reduce to the special  case of \sy\ for
comparison with earlier results, as described above.  Finally in the
Appendices we explain some of the group theory  involved in the
calculations  and give explicit results for $SO(N)$ and $Sp(N)$ for the
one-loop  \epscalar{} quartic interaction $\beta$-functions.

\section{\label{sec::rev}Gauge theory with fermions}

Consider a non-abelian gauge theory with gauge fields $W_{\mu}^a$ 
and a multiplet of  
two-component fermions $\psi_{\alpha}^{A}(x)$ transforming according to  a 
representation $R$ of the gauge group ${\cal G}$. 

The Lagrangian density (in terms of bare fields) is 
\be
L_B = -\frak{1}{4}G^2_{\mu\nu} - \frak{1}{2\alpha}(\pa^{\mu}W_{\mu})^2 + 
C^{a*}\pa^{\mu}D_{\mu}^{ab}C^b + 
i\psib_{\alphadot A}\sigb^{\mu\alphadot\alpha}(D_{\mu})^{A}{}_B\psi^{B}_{\alpha} 
\ee
where 
\be
G^a_{\mu\nu} = \pa_{\mu}W_{\nu}^a - \pa_{\nu}W_{\mu}^a 
+ gf^{abc}W_{\mu}^b W_{\nu}^c
\ee
and 
\be
(D_{\mu})^{A}{}_B = \delta^{A}{}_B\pa_{\mu} 
- ig (R^a )^{A}{}_BW_{\mu}^a
\ee
and the usual covariant gauge fixing and ghost $(C, C^*)$ terms have 
been introduced. As usual $\sigb^{\mu} \equiv (I, -\hbox{\boldmath$\sigma$})$ 
where {\boldmath$\sigma$} 
are the Pauli matrices. 

For the case when the theory admits a gauge invariant fermion mass 
term we will have $L_B \to L_B + L_B^m$, where 
\be
L_B^m = \frak{1}{2} m_{AB}\psi^{\alpha A}\psi^B_{\alpha} 
+ \hbox{c.c.} 
\ee

Dimensional reduction amounts to  imposing that all field variables
depend only  on a subset of the total number of space-time dimensions;
in this case $d$ out of $4$ where  $d = 4 - 2\ep$. We can then make the
decomposition 
\be
W_{\mu}^a(x^j ) = \{ W_i^a (x^j ), W_{\sigma}^a(x^j )\}
\ee
where
\be
\delta^i{}_i = \delta^j{}_j = d \qq \hbox{and}\qq \delta_{\sigma\sigma} = 2\ep.
\ee
It is then easy to show that
\be
L_B = L _B^d + L_B^{\ep} 
\ee
where
\be
L _B^d = -\frak{1}{4}G^2_{ij} -\frak{1}{2\alpha}(\pa^{i}W_{i})^2 +
C^{*}\pa^{i}D_{i}C + 
i\psib\sigb^i D_{i}\psi 
\label{eq:AD}
\ee
and 
\be
 L_B^{\ep} = \frak{1}{2}(D_i W_{\sigma})^2 
- g\psib\sigb_{\sigma}R^a\psi W_{\sigma}^a
-\frak{1}{4}g^2 f^{abc}f^{ade}W^b_{\sigma}W^c_{\sigp}W^d_{\sigma}W^e_{\sigp}.
\label{eq:AE}
\ee
Conventional dimensional regularisation (\dreg) amounts to using 
Eq.~(\ref{eq:AD}) and discarding 
Eq.~(\ref{eq:AE}).

We would now like to rewrite Eq.~(\ref{eq:AD}) and Eq.~(\ref{eq:AE}) in
terms of renormalised quantities. It is clear, however, from
the dimensionally reduced form of the gauge transformations: 
\besub
\bea 
\delta W^a_i &=& \pa_i\Lambda^a + gf^{abc}W^b_i\Lambda^c\label{eq:AFa}\\
 \delta W^a_{\sigma}&=& gf^{abc}W^b_{\sigma}\Lambda^c\label{eq:AFb}\\ 
\delta\psi^{A} &=& ig(R^a)^{A}{}_B \psi^B\Lambda^a 
\label{eq:AFc} 
\eea
\eesub 
that each term in
Eq.~(\ref{eq:AE}) is separately invariant under gauge  transformations.
The $W_{\sigma}$-fields behave exactly like scalar  fields, and are hence
known as \epscalars.  There is
therefore no  reason to expect the $\psib\psi W_{\sigma}$ vertex to
renormalise in the same  way as the $\psib\psi W_i$ vertex (except in
the case of supersymmetric theories). In the  case of the quartic
\epscalar{} interaction it is evident that  
more than one such coupling is permitted by Eq.~(\ref{eq:AFb}).
In  other words, we cannot in general expect the $f-f$~tensor structure 
present in Eq.~(\ref{eq:AE}) to be preserved under renormalisation.
This is clear from the abelian  case, where there is no quartic
interaction in $L_B^{\ep}$ but there is a divergent graph at one
loop from a fermion loop.

We are therefore led to consider the following expressions
\footnote{Since \epscalars{} are present only on internal lines 
we could, in fact, choose the
wave function renormalisation of $W_{\sigma}$ and $W_i$ to be the same; or,
indeed, have no wave function renormalisation for $W_{\sigma}$ at all. The
crucial thing is correct treatment of sub-graphs, which means recognition that
vertices with \epscalars{} renormalise in a different way from their gauge
counterparts. However, we choose to renormalise the \epscalar{} 
conventionally.} for renormalised quantities 
$L^d$ and $L^{\varepsilon}$ : 
\bea
L^d = &-& \frak{1}{4}Z^{WW} (\pa_i W_j - \pa_j W_i )^2
-\frak{1}{2\alpha}(\pa^{i}W_{i})^2\nn
&-& Z^{WWW}gf^{abc}\pa_i W_j^a W^{bi} W^{cj} -
\frak{1}{4}Z^{4W}g^2f^{abc}f^{ade}W_i^b W_j^cW^{di} W^{ej}\nn
&+& Z^{C C}\pa^i C^*\pa_i C + Z^{CCW}gf^{abc}\pa^i C^{a*}W_i^b C^c\nn
&+& Z^{\psi\psi}i\psib\sigb^i\pa_i\psi + Z^{\psi\psi W}g\psib R^a\sigb^i\psi W^a_i
\label{eq:AFA}
\eea
 and
\bea
L^{\varepsilon} & =&\frak{1}{2}Z^{\varepsilon\varepsilon}(\pa_i W_{\sigma})^2
+ Z^{\varepsilon\varepsilon W} gf^{abc}\pa_i W_{\sigma}^a W^{bi} W_{\sigma}^c\nn
&+& Z^{\varepsilon\varepsilon WW}g^2f^{abc}f^{ade}W_i^b W_{\sigma}^c 
W^{di} W_{\sigma}^e
-Z^{\psi\psi\varepsilon}g_e\psib R^a\sigb_{\sigma}\psi W^a_{\sigma}\nn
           &-& \frak{1}{4} \sum_{r=1}^p Z_r^{4\varepsilon}\la_r
            H^{abcd}_rW^a_{\sigma}W^c_{\sigp}W^b_{\sigma}W^d_{\sigp}.
\label{eq:AFB}
\eea
In the case when we have a fermion mass term we would also have
\be
L^m = \frak{1}{2}Z_m Z^{\psi\psi}m\psi^{\alpha}\psi_{\alpha} + \hbox{c.c.} 
\ee

Eq.~(\ref{eq:AFA}) is the usual expression for the Lagrangian in terms
of  renormalised parameters. In Eq.~(\ref{eq:AFB}) we have introduced a 
``Yukawa'' coupling $g_e$ and a set of $p$ 
quartic couplings $\la_r$. (Strictly speaking, \eqn{eq:AFB}
should also have a mass term for the \epscalars; but since 
this mass term does not affect $\beta_s$ or $\gamma_m$ we omit it here.) 
The number $p$ is given  by the number of independent rank four tensors
$H^{abcd}$ which  are non-vanishing when symmetrised with respect to
$(ab)$ and $(cd)$ interchange. In the next section we discuss the 
quartic vertex and its renormalisation in more detail.

\section{\label{sec::epsfour}The \epscalar{} self coupling}

Let us discuss the structure  of the quartic \epscalar{} couplings
for an  arbitrary gauge group. These interactions are  invariant under the 
symmetry ${\cal G}\otimes O(2\epsilon)$, where only the  ${\cal G}$ is 
gauged. The renormalisation properties of 
scalar theories with invariances of the type $G_1\otimes G_2$ have been 
studied in considerable detail, for example $O(m)\otimes O(n)$ 
in the theory of critical phenomena and $U(m)\otimes U(n)$ in the 
context of \qcd. In these cases, however, the scalars transform 
as vector (fundamental) representations of the gauge group factors 
whereas for us they transform as adjoints.

This raises an interesting group theory
question: how  many independent couplings are there for a given gauge
group ${\cal G}$? Evidently  the question of how many independent 
tensors of the form $K^{abcd}$ there are is the question  of how many
times the singlet representation occurs in the reduction to  irreducible
representations of the product of four adjoint representations. 
(Neither this question, nor the obvious generalisation 
to $n$-tensors $K^{a_1 \cdots a_n}$  
has been  much studied in the literature; an exception being 
the classic work of Cvitanovic~\cite{predrag}, to which we will return presently). 
The set 
of tensors relevant to our problem is the subset of such tensors 
$H^{abcd}$ which 
is invariant with respect to $(a,b)$ and $(c,d)$ exchange, 
because of the $O(2\epsilon)$
invariance. 

If we have an irreducible basis of dimensionality $\gamma(n)$ 
for the $n$-tensors of the form  
\be K_{\alpha}^{a_1 \cdots a_n}, 
\quad 1 \leq \alpha \leq \gamma(n) 
\label{eq:baseone}
\ee  then a general $n$-tensor $K^{a_1\cdots a_n}$ can be
expressed in terms of the basis  as  
\be K^{a_1\cdots a_n} = x_{\beta}K_{\beta}^{a_1\cdots a_n}, \ee 
where $x_{\beta}$ are determined by the
equation   
\be Q^n_{\alpha\beta}x_{\beta} = y_{\alpha} 
= K^{a_1\cdots a_n}K_{\alpha}^{a_1 \cdots a_n}
\ee
and  
\be
Q^n_{\alpha\beta} = K_{\alpha}^{a_1\cdots a_n} K_{\beta}^{a_1\cdots a_n}.
\label{eq:basetwo}
\ee
Thus construction of the $Q^n$-matrix permits reduction of an arbitrary 
$n$-tensor to the basis. 

\subsection{The case ${\cal G} = SU(N)$}

The fundamental representation $T^a$ of the generators $R^a$ of 
$SU(N)$ satisfies
\bea
\left[T^a, T^b \right] &=& if^{abc}T^c\nn
\left\{T^a, T^b \right\} &=& d^{abc}T^c + \frac{b}{N}\delta^{ab}\nn
\Tr\left(T^a T^b\right) &=& \frac{b}{2}\delta^{ab},
\label{eq:sunalg}
\eea
where $b$ is a constant. For the rest of this section 
we will adopt the usual convention 
whereby $b = 1$.

In table~\ref{gamsun}\ we present some results for the dimensionality $\gamma (n)$
for $SU(N)$ as a function of $N$. It is interesting that 
Cvitanovic~\cite{predrag}\ remarks 
that a formula for the dimensionality of a  basis  
(in general over-complete) is provided by the subfactorial 
$\beta(n)$ where 
\be
\beta(n) = n! ( 1 - \frac{1}{1!}+\frac{1}{2!} + \ldots (-1)^n \frac{1}{n!}). 
\ee
\begin{table}
\begin{center}
\begin{tabular}{|c| c c c c c|} \hline
$n$ & $N=2$ & $N=3$ 
& $N=4$ & $N=5$ & $N=6$ \\ \hline
& & & & & \\ 
$2$ & $1$  & $1$ 
& $1$ & $1$ & $1$ \\ 
& & & & & \\ \hline
& & & & & \\ 
$3$ & $1$  & $2$ 
& $2$ & $2$ & $2$ \\ 
& & & & & \\ \hline
& & & & & \\ 
$4$ & $3$  & $8$ 
& $9$ & $9$ & $9$ \\ 
& & & & & \\ \hline
& & & & & \\ 
$5$ & $\cdots$  & $\cdots$ 
&$\cdots$ & $\cdots$ & $\cdots$ \\ 
& & & & & \\ \hline
\end{tabular}
\caption{\label{gamsun}Basis dimensionality for $n$-tensors in $SU(N)$.}
\end{center}
\end{table}
It appears that for sufficiently large $N$ we have $\gamma (n) = \beta(n)$. 

A natural choice for the basis for the case $n=4$ when $N \geq 4$ 
is  
given by\footnote{An alternative way to define a basis which has the virtue of 
being immediately generalisable to any group~\cite{predrag} is in terms of traces of products 
of the generators in the defining representation, thus 
$\Tr\left(T^a T^b T^c T^d\right)$, 
$\Tr\left(T^a T^b\right)\Tr\left(T^c T^d\right)$
 etc.} 
\bea
K_1 &=& \delta^{ab}\delta^{cd}
\quad K_4 = d^{abe}d^{cde}\quad K_7 = d^{abe}f^{cde}\nn  
K_2 &=& \delta^{ac}\delta^{bd}
\quad K_5 = d^{ace}d^{bde}\quad K_8 = d^{ace}f^{bde}\nn  
K_3 &=& \delta^{ad}\delta^{bc}
\quad K_6 = d^{ade}d^{bde}\quad K_9 = d^{ade}f^{bce}.
\eea
The reduction of the basis to $\gamma=8$ in the case $SU(3)$ is 
achieved via the relation~\cite{msw,Dittner:1972hm} 
\be
K_4 + K_5 + K_6 = \frac{1}{3}(K_1 + K_2 + K_3)
\ee
which is not valid for $N \geq 4$. The corresponding identity for general $N$ 
reduces a symmetrised $(N+1)$-tensor consisting of $N-1$ $d$-tensors;    
for an elegant derivation see \reference{rashid}.

For the \epscalar{} interactions a possible basis for $N \geq 4$ is therefore 
\bea
H_1 &=& \frak{1}{2}K_1\nn
H_2 &=& \frak{1}{2}(K_2 + K_3)\nn
H_3 &=& \frak{1}{2}K_4\nn
H_4 &=& \frak{1}{2}(K_5 + K_6).
\eea
Note that the absence of a $d-f$ type term from the basis follows from the
identity
\be
K_8 + K_9 = -f^{abe} d^{cde}.
\ee
Let us introduce the couplings
\begin{eqnarray}
  \alpha_s = \frac{g^2}{4\pi}\,,
  \quad
  \alpha_e = \frac{g_e^2 }{4\pi}
  \quad\mbox{and}\quad 
  u_r = \frac{\lambda_r}{4\pi}
  \,,
\end{eqnarray}
and define the corresponding $\beta$ functions for the $u_r$ couplings
\be
{\beta_{u_r}
   = \mu^2 \frac{{\rm d}}{{\rm d}\mu^2} \frac{u_r}{\pi}}.
\ee
If we write (with the normalisation of Eq.~(\ref{eq:AFB}))
\be
\la_r H^{abcd}_r \to  \sum_{r=1}^4 4\pi  u_r H_r
\ee
then the $\beta$-functions for the $u_r$ couplings
 are given at one loop by\footnote{Here and for the rest of this 
section we  
suppress a factor of $1/\lfp$ in every one-loop $\beta$-function.}
\bea
\beta_{u_1} &=& 8u_1^2 + 4N^2u_1u_2+12u_2^2\nn
&+& \frac{4(N^2-4)}{N}\left\{u_1u_3+u_1u_4+2u_2u_4 
+ \frac{1}{N}(u_3^2+2u_3u_4+3u_4^2)\right\}\nn
\beta_{u_2} &=& 12u_1u_2+ (2N^2+6)u_2^2\nn
&+& \frac{2(N^2-4)}{N}\left\{2u_2u_3+2u_2u_4 
+ \frac{1}{N}(u_3^2+6u_3u_4+3u_4^2)\right\}\nn
\beta_{u_3} &=& 12u_1u_3+ 4u_2u_3+16u_2u_4\nn
&+& \frac{1}{N}\left\{(3N^2-40)u_3^2+6(N^2-12)u_3u_4 +(7N^2-96)u_4^2\right\}\nn
\beta_{u_4} &=& 12u_1u_4+ 8u_2u_3+12u_2u_4\nn
&+& \frac{1}{N}\left\{4(N^2-14)u_4^2+4(N^2-18)u_3u_4 -8u_3^2\right\}
\label{eq:betau}
\eea
where for the moment we suppress contributions from the gauge coupling $\alpha_s$ and  
\epscalar{} Yukawa coupling $\alpha_e $.

Because of   the nature of the bare theory, and to explore more easily the 
supersymmetric case, it is natural to consider alternative bases, for example:
\bea
\Hbar_1 &=& H_1\nn
\Hbar_2 &=& H_2\nn
\Hbar_3 &=& \frak{1}{2}(f^{ace}f^{bde} + f^{ade} f^{bce})\nn
\Hbar_4 &=& \frak{1}{2}(f^{aef}f^{bfg}f^{cgh}f^{dhe}
+f^{aef}f^{bfg}f^{dgh}f^{che}).
\label{eq:hbarbasis}
\eea
We shall also see that it is this kind of basis (avoiding 
use of the $d$-tensor) that generalises most easily to other groups.
 
We have 
\bea
\Hbar_3 &=& \frac{4}{N}H_1 - \frac{2}{N}H_2 + 2 H_3 - H_4\nn
\Hbar_4 &=& 2H_1 +H_2 + \frac{N}{2}H_3, 
\eea
so that if we write 
\be
\sum_{r=1}^4 u_r H_r =  \sum_{r=1}^4 v_r \Hbar_r
\ee
then
\bea
v_1 &=& u_1 -\frac{4}{N}(u_3 + u_4)\nn 
v_2 &=& u_2 -\frac{2}{N}u_3 -\frac{6}{N}u_4\nn
v_3 &=& - u_4\nn
v_4 &=& \frac{2}{N}u_3 + \frac{4}{N} u_4.
\eea
The $\beta$-functions for the $v_r$ couplings are given at one loop by 
\bea
\beta_{v_1} &=& 8v_1^2 + 4N^2v_1v_2+12v_2^2
- 4Nv_1v_3+6N^2v_1v_4+8Nv_2v_3\nn
&+&8Nv_3v_4 + 8N^2 v_2v_4 +10 N^2 v_4^2 -12N  v_1 \as \nn
\beta_{v_2} &=& 12v_1v_2+ (2N^2+6)v_2^2 -4Nv_2v_3+ 6N^2 v_2v_4\nn
&-&4Nv_3v_4+3N^2 v_4^2-12 N v_2 \as \nn
\beta_{v_3} &=& 4Nv_3^2+12v_1v_3- 4v_2v_3 -4Nv_2v_4 + (2N^2+8)v_3v_4\nn
&-&2Nv_4^2 -12Nv_3 \as \nn
\beta_{v_4} &=& (\frak{3}{2}N^2 + 16)v_4^2 + 12v_1v_4 + 20v_2v_4-2v_3^2\nn
&-&2Nv_3v_4-12Nv_4\as +6\as^2 
\label{eq:betav}
\eea
where we have now included the gauge coupling contribution (note that 
the $\as^2 $ terms contribute only to $\beta_{v_4}$).

Another choice of basis (in fact the one employed in 
Ref.~\cite{Harlander:2006rj}) is 
\bea
\Htil_1 &=&  \delta^{ac}\delta^{bd} +  \delta^{ad}\delta^{bc} 
+  \delta^{ab}\delta^{cd}\nn
\Htil_2 &=& \frak{1}{2}( \delta^{ac}\delta^{bd} +  \delta^{ad}\delta^{bc})
-  \delta^{ab}\delta^{cd}\nn
\Htil_3 &=& \Hbar_3\nn
\Htil_4 &=& \Hbar_4
\eea
so that 
\bea
\Htil_1 &=& 2(\Hbar_1 + \Hbar_2)\nn
\Htil_2 &=&  \Hbar_2 - 2\Hbar_1
\eea
and if we write 
\be
\sum_{r=1}^4 v_r \Hbar_r =  \sum_{r=1}^4 w_r \Htil_r 
\ee
then
\bea
w_1 &=& \frak{1}{6}(v_1 + 2v_2)\nn 
w_2 &=& \frak{1}{3}(-v_1 + v_2)\nn
w_3 &=& v_3\nn
w_4 &=& v_4.
\eea
In this basis the $\beta$-functions become
\bea
\beta_{w_1} &=& \frak{1}{3}[
(112+16N^2)w_1^2+(4N^2-8)w_1w_2-4Nw_1w_3+26N^2w_1w_4\nn
&-&(2N^2-4)w_2^2
+4Nw_2w_3+4N^2w_2w_4+8N^2w_4^2] -12N  w_1 \as \nn
\beta_{w_2} &=& \frak{1}{3}[
-8(N^2+1)w_1^2+16(N^2+1)w_1w_2-16Nw_1w_3-16N^2w_1w_4\nn &+&(10N^2-62)w_2^2
-20Nw_2w_3+10N^2w_2w_4-12Nw_3w_4-7N^2w_4^2]\nn  &-&12N  w_2 \as \nn
\beta_{w_3} &=& 16w_1w_3-8Nw_1w_4-28w_2w_3-4Nw_2w_4+4Nw_3^2\nn 
&+& (2N^2 +8)w_3w_4 -2Nw_4^2 -12N  w_3 \as \nn
\beta_{w_4} &=& 64w_1w_4-4w_2w_4-2w_3^2-2Nw_3w_4 + 
(\frak{3}{2}N^2 +16)w_4^2\nn
&-&12N w_4 \as  +6\as^2 .
\label{eq:betaw}
\eea
For the rest of the paper we will use the $v$-basis; 
as already remarked the use of $\Hbar_{3,4}$ means the generalisation 
to other groups can be carried out easily.

\subsubsection{The fermion contribution}

The contribution of the fermion loop to the scalar anomalous dimension
results  in a contribution of  
\be
\Delta \beta_{u_i} = 8 n_f I_2 (R) \aep  u_i
\label{eq:psiloop}
\ee
to each $\beta$-function  in Eq.~(\ref{eq:betau}), with corresponding
contributions to  Eq.~(\ref{eq:betav}) and Eq.~(\ref{eq:betaw}).

In \eqn{eq:psiloop}\ and subsequently we follow the following convention: 
our fermion representation consists of 
$n_f$ sets of Dirac fermions or $2n_f$ sets of two-component fermions, in 
irreducible representations with identical Casimirs; and the whole 
representation must of course be anomaly free. 
We will pay 
particular attention to the case of an adjoint representation 
with $n_f = \frac{1}{2}$, which is supersymmetric, and to  
the case of $n_f$ flavours, that is $n_f$ sets of fundamental
two component fermions with $n_f$ sets of anti-fundamental
two component fermions, which is \qcd. For the definition of $I_2 (R)$ and 
more details on group theoretic considerations see Appendix~\ref{appA}.

The 1PI
fermion box diagram  makes a contribution to the $\beta$-functions
(appropriately normalised)  of the form
\bea
\Hbar_i\Delta\beta_{v_i} &=& -4n_f \aep^2  \bigl[
\Tr(R^a R^b R^c R^d) +\Tr(R^a R^b R^d R^c) +\Tr(R^a R^d R^c R^b)\nn 
&+&\Tr(R^a R^c R^d R^b)-\Tr(R^a R^d R^b R^c) -\Tr(R^a R^c R^b R^d)\bigr].
\label{eq:floop}
\eea 
For a general representation this is not easily expressed in terms of 
one of our choice of bases. In the special case of  an adjoint
representation (with $n_f = \frac{1}{2}$), we  find that
\be
\Hbar_i\Delta\beta_{v_i} =  \aep^2  (-2 N \Hbar_3 -4 \Hbar_4) 
\label{eq:fadj}
\ee 
so that the complete set of $\beta$-functions for the case of an $SU(N)$ 
gauge theory with an adjoint fermion multiplet is: 
\bea
\beta_{v_1} &=& 8v_1^2 + 4N^2v_1v_2+12v_2^2
- 4Nv_1v_3+6N^2v_1v_4+8Nv_2v_3 \nn
&+& 8Nv_3v_4 + 8N^2 v_2v_4 +10 N^2 v_4^2 -12N  v_1 \as  + 4Nv_1\aep \nn
\beta_{v_2} &=& 12v_1v_2+ (2N^2+6)v_2^2 -4Nv_2v_3+ 6N^2 v_2v_4\nn
&-& 4Nv_3v_4+3N^2 v_4^2-12 N v_2 \as  + 4Nv_2\aep \nn
\beta_{v_3} &=& 4Nv_3^2+12v_1v_3- 4v_2v_3 -4Nv_2v_4 + (2N^2+8)v_3v_4\nn
&-&2Nv_4^2 -12Nv_3 \as + 4Nv_3\aep  -2N \aep^2 \nn
\beta_{v_4} &=& (\frak{3}{2}N^2 + 16)v_4^2 + 12v_1v_4 + 20v_2v_4-2v_3^2\nn
&-& 2Nv_3v_4-12Nv_4\as +6\as^2  + 4Nv_4\aep  -4 \aep^2 .
\label{eq:betavtota}
\eea
If we now set $v_1 = v_2 = v_4 =0$ and $v_3 = \aep  = \as $ the theory becomes 
supersymmetric; and substituting these values in   Eq.~(\ref{eq:betavtota}) we 
indeed find $\beta_{v_1} = \beta_{v_2} = \beta_{v_4} =0$ and 
\be
\beta_{v_3} = -\frac{6N}{8\pi^2}\as^2, 
\ee
(restoring the $8\pi^2$ factor) which is identical to the 
one-loop gauge $\beta$-function $\beta_s$ in the \sic\ case.  

Let us consider now the special case of $SU(3)$. 
In $SU(3)$, $(\Hbar_1, \Hbar_2, \Hbar_3)$ form a basis; however  if we set 
$v_4 = 0$ in Eq.~(\ref{eq:betavtota}) then we nevertheless have 
\be
\beta_{v_4} = -2v_3^2 +6\as^2  -4\aep^2 .
\ee
This represents a set of contributions to $\beta_{v_{1,2,3}}$ which we 
can identify by using the identity
\be
\Hbar_4 = \frak{3}{2}(\Hbar_1 + \Hbar_2) + \frak{1}{2}\Hbar_3.
\ee

Incorporating these contributions into Eq.~(\ref{eq:betavtota})\  we 
thus find for in $SU(3)$ 
\bea
\beta_{v_1} &=& 8v_1^2 + 36v_1v_2+12v_2^2 -12 v_1 v_3+24v_2v_3 -36  v_1
\as \nn  
&+& 12v_1\aep   -3v_3^2 + 9\as^2  -6\aep^2 \nn
\beta_{v_2} &=& 12v_1v_2+ 24v_2^2 -12v_2v_3 -36 v_2 \as \nn 
&+& 12v_2\aep  -3v_3^2 + 9\as^2  -6\aep^2 \nn
\beta_{v_3} &=& 11v_3^2+12v_1v_3- 4v_2v_3  -36v_3 \as \nn 
&+& 12v_3\aep    + 3\as^2 -8\aep^2. 
\label{eq:betavtot3a}
\eea
It is easy to check that this set still reduces 
correctly in the supersymmetric limit.

For the special case of $SU(2)$, the basis is two dimensional and 
we have the identities
\bea
2\Hbar_1 -\Hbar_2 -\Hbar_3 &=& 0\nn
2\Hbar_1 +\Hbar_2 -\Hbar_4 &=& 0.
\label{eq:hbarsu2}
\eea
If we choose the basis $(\Hbar_1,\Hbar_2)$ then we find 
\bea
\beta_{v_1} &=& 8v_1^2 + 16v_1v_2+12v_2^2 -24 v_1 \as  + 8v_1\aep 
-16\aep^2 +12\as^2 \nn
\beta_{v_2} &=& 12v_1v_2+ 14v_2^2  -24 v_2 \as  + 8v_2\aep  +6\as^2. 
\eea
Alternatively we could choose the basis $(\Hbar_3,\Hbar_4)$
when we find
\bea
\beta_{v_3} &=& 8v_3^2 + 24v_3v_4 - 24v_3 \as +8v_3\aep -4 \aep^2  \nn
\beta_{v_4} &=& 38v_4^2 -2v_3^2 -4v_3v_4-24v_4\as +6\as^2  + 8v_4\aep  
-4 \aep^2 .
\label{eq:betavtota2}
\eea
With this basis the \sic\ limit is again apparent; setting 
$v_4 = 0$ and $v_3 = \aep  = \as $ we obtain 
$\beta_{v_4} = 0$ and $\beta_{v_3} = -12\as^2 $ as expected.

For the fundamental representation of $SU(N)$ 
we find 
\be
\Tr(R^a R^b R^c R^d) = \frac{1}{4N} \left[
K_1 - K_2 +K_3\right] 
+ \frac{1}{8}\left[K_4 -K_5 + K_6\right]
+\frac{i}{8}\left[K_7 + K_8 + K_9\right] 
\label{eq:tracesu}
\ee
and hence for the case of $2n_f$ sets of fermions in the 
fundamental representation of $SU(N)$ 
(corresponding to \qcd{} with $n_f$ flavours), 
\be
\Hbar_i\Delta\beta_{v_i} 
= 2n_f\aep^2 \left[\frac{2}{N}(\Hbar_1 +\Hbar_2 -\Hbar_4)
-\Hbar_3\right]
\label{eq:suferms}
\ee
so that 
the complete set of $\beta$-functions for this case is: 
\bea
\beta_{v_1} &=& 8v_1^2 + 4N^2v_1v_2+12v_2^2
- 4Nv_1v_3+6N^2v_1v_4+8Nv_2v_3 \nn
&+& 8Nv_3v_4 + 8N^2 v_2v_4 +10 N^2 v_4^2 -12N  v_1 \as  
+ 4n_fv_1\aep  +4\frac{n_f}{N}\aep^2 \nn
\beta_{v_2} &=& 12v_1v_2+ (2N^2+6)v_2^2 -4Nv_2v_3+ 6N^2 v_2v_4\nn
&-& 4Nv_3v_4+3N^2 v_4^2-12 N v_2 \as  + 4n_fv_2\aep +4\frac{n_f}{N}\aep^2 \nn
\beta_{v_3} &=& 4Nv_3^2+12v_1v_3- 4v_2v_3 -4Nv_2v_4 + (2N^2+8)v_3v_4\nn
&-&2Nv_4^2 -12Nv_3 \as + 4n_fv_3\aep  -2n_f \aep^2 \nn
\beta_{v_4} &=& (\frak{3}{2}N^2 + 16)v_4^2 + 12v_1v_4 + 20v_2v_4
-2v_3^2\nn
&-& 2Nv_3v_4-12N v_4\as +6\as^2  + 4n_fv_4\aep -4\frac{n_f}{N}\aep^2 .
\label{eq:betavtotf}
\eea
It is straightforward to incorporate the fermion contributions in 
our other choices of bases involving $u_i$ or $w_i$.

Turning again to the special case of $SU(3)$, and setting  
$v_4 = 0$ in Eq.~(\ref{eq:betavtotf}) we have 
\be
\beta_{v_4} = -2v_3^2 +6\as^2  -4\frac{n_f}{3}\aep^2 
\ee
and incorporating these contributions into Eq.~(\ref{eq:betavtotf})\  we 
thus find for $SU(3)$: 
\bea
\beta_{v_1} &=& 8v_1^2 + 36v_1v_2+12v_2^2 -12v_1 v_3 
+24v_2v_3 -36  v_1 \as \nn  
&+& 4n_fv_1\aep   -3v_3^2 + 9\as^2  -2\frac{n_f}{3}\aep^2 \nn
\beta_{v_2} &=& 12v_1v_2+ 24v_2^2 -12v_2v_3 -36 v_2 \as \nn 
&+& 4n_fv_2\aep  -3v_3^2 + 9\as^2  -2\frac{n_f}{3}\aep^2 \nn
\beta_{v_3} &=& 11v_3^2+12v_1v_3- 4v_2v_3  -36v_3 \as \nn 
&+& 4n_fv_3\aep   + 3\as^2 -8\frac{n_f}{3} \aep^2. 
\label{eq:betavtot3f}
\eea
The special case of $SU(2)$ in the fundamental fermion case we leave 
as an exercise for the reader.

\subsection{The general case}
In this subsection we give the results for $\beta_{v_i}$ 
for a general gauge group. The various group invariants are defined 
in Appendix~\ref{appA}, where results for them for the fundamental 
representations of $SU(N)$, $SO(N)$ 
and $Sp(N)$ also appear. 

We have derived these results both by substituting in the general expressions 
that follow and by direct calculations with each class of group 
in the manner described in the previous section.


We find
\bea
\beta_{v_1}&=&-32 n_f\frac{5C_A^2D_2 (RA) 
+ (C_A - 6C_R)D_2 (A)I_2(R)}
  {25C_A^4N_A - 12D_2 (A)(2 + N_A)}\aep^2 
- 12C_Av_1 \as \nn
&+& 8I_2(R)n_fv_1 \aep 
+ 
 8v_1^2 + 12v_2^2 - \frac{192D_2 (A) - 80C_A^4N_A}
  {9C_AN_A(N_A-3)}v_3v_4 \nn
&+& 
 \frac{4}{27N_A}
\Bigl\{7\frac{-12D_2 (A) + 5C_A^4N_A}{N_A-3} \nn
&-&
   24 \frac{72 D_2 (A)^2 
- 90C_A^2D_3 (A)N_A
+ 25C_A^4D_2 (A)N_A}
     {25C_A^4N_A - 12D_2 (A)(2 + N_A)}\Bigr\}v_4^2 \nn
&+& v_1[4(1 + N_A)v_2 - 4C_Av_3 + 6C_A^2v_4] 
+ v_2(8C_Av_3 + 8C_A^2v_4)\nn
\beta_{v_2}&=&-32 n_f\frac{5C_A^2D_2 (RA) 
+ (C_A - 6C_R)D_2 (A)I_2(R)}{
  25C_A^4N_A - 12D_2 (A)(2 + N_A)} \aep^2\nn
&-& 12C_Av_2\as  + 8I_2(R)n_fv_2\aep  + 
 12v_1v_2 + 2(4 + N_A)v_2^2 \nn
&+& \frac{96D_2 (A) - 40C_A^4N_A}{9C_AN_A(N_A-3)}v_3v_4 
+ \frac{2}{27N_A}\Bigl\{7\frac{12D_2 (A) 
- 5C_A^4N_A}{(N_A-3)} \nn
&-& 48\frac{72 D_2 (A)^2 
- 90C_A^2D_3 (A)N_A 
+ 25C_A^4D_2 (A)N_A}{
     25C_A^4N_A - 12D_2 (A)(2 + N_A)}\Bigr\}v_4^2 \nn
&+& 
          v_2(-4C_Av_3 + 6C_A^2v_4)\nn
\beta_{v_3}&=&-\frac{4 n_f}{25C_A^4N_A 
-12D_2 (A)(2 + N_A)}\Bigl\{
35C_A^4I_2(R)N_A - 10C_A^3C_RI_2(R)N_A\nn
&+& 4C_AD_2 (RA)(2 + N_A) -  
    16D_2 (A)I_2(R)(2 + N_A)\Bigr\}\aep^2 \nn
&-& 
 12 C_Av_3 \as+ 8I_2(R)n_fv_3\aep  + 12v_1v_3 + 4C_Av_3^2 \nn
&+& 
 2\frac{48D_2 (A)(-1 + N_A) + C_A^4N_A(-61 + 7N_A)}
  {9C_A^2N_A(N_A-3)}v_3v_4 \nn
&-& 
 \frac{4}{
  27C_A(N_A-3)N_A[25C_A^4N_A - 12D_2 (A)(2 + N_A)]}\nn
&&\Bigl\{144 D_2 (A)^2(2 + N_A)(1 + 2N_A) \nn
&+& 12C_A^4D_2 (A)N_A
     [-191 + (-56 + N_A)N_A] \nn
&+& C_A^2N_A[-216D_3 (A)(-3 + N_A)
       (2 + N_A) + 25C_A^6N_A(23 + 4N_A)]\Bigr\}v_4^2 \nn
&-& 
         v_2(4v_3 + 4C_Av_4)\nn
\beta_{v_4}&=&6\as^2  + 8 n_f\frac{5C_A^2(C_A - 6C_R)I_2(R)N_A 
         + 12D_2 (RA)(2 + N_A)}{25C_A^4N_A 
- 12D_2 (A)(2 + N_A)}\aep^2\nn
&-& 2v_3^2 - 12C_Av_4\as  + 8 I_2(R)n_fv_4 \aep 
         + 12v_1v_4 + 20v_2v_4 - 2C_Av_3v_4 \nn
&-&
         \frac{1152D_3 (A)(2 + N_A) 
-5C_A^2[125C_A^4N_A + 
            4D_2 (A)(98 + N_A)]}{6[25C_A^4N_A 
            - 12D_2 (A)(2 + N_A)]}v_4^2.
\label{eq:genbetavs}
\eea
The forms taken by $C_{A,R}$, $I_2(R)$ and the various invariants 
$D_2(A)$ etc for $SU(N)$, $SO(N)$ and $Sp(N)$ are given in 
Tables~\ref{gaminvsa}-\ref{gaminvsc}\ in the Appendix. 
Using Table~\ref{gaminvsa}, it is easy to show that the results in 
Eq.~(\ref{eq:genbetavs})
reduce to the results in Eq.~(\ref{eq:betavtotf}) for the case of $SU(N)$.

\section{\label{sec::betafour}The general four-loop results}

The renormalisation constants for the various couplings
are defined through
\begin{align}
  g_s^0   &= \mu^{\epsilon}Z_s g_s\,,\qquad & 
g_e^0   &= \mu^{\epsilon}Z_e g_e\,,\qquad &
  \sqrt{v_r^0} &= \mu^\epsilon Z_{v_r} \sqrt{v_r}\,,
  \nonumber\\
  \varepsilon^{0,a}_\sigma &= \sqrt{Z^{\varepsilon\varepsilon}} \varepsilon^a_\sigma\,,
  \qquad &
  \Gamma_{\psib\psi\varepsilon}^{0} &= 
  Z^{\psi\psi\varepsilon} \Gamma_{\psib\psi\varepsilon}\,,\qquad &
  \Gamma^{r,0}_{\varepsilon\varepsilon\varepsilon\varepsilon}
  &= Z^{4\varepsilon}_r \Gamma^{r}_{\varepsilon\varepsilon\varepsilon\varepsilon}
  \,,
  \label{eq::renconst}
\end{align}
where $\Gamma_{\psib\psi\varepsilon}$  and
$\Gamma_{\varepsilon\varepsilon\varepsilon\varepsilon}$ are the
one-particle irreducible \epscalar{}--fermion and
four-\epscalar{} Green functions, respectively, the superscript 
``0'' denotes bare quantities, and $\mu$ is the renormalisation scale.
The renormalisation constants associated with the various couplings 
satisfy the following
relations
\begin{equation}
\begin{split}
  Z_s &= \frac{Z^{\psi\psi W }}{Z^{\psi\psi}\sqrt{Z^{WW}}}\,,\qquad
  Z_e = \frac{Z^{\psi\psi\varepsilon}}
{Z^{\psi\psi}\sqrt{Z^{\varepsilon\varepsilon}}}
  \,,\qquad
  Z_{v_r} = \frac{\sqrt{Z^{4\varepsilon}_r}}
{Z^{\varepsilon\varepsilon}}
  \,,
\end{split}
\end{equation}
with renormalisation constants as defined in \eqn{eq:AFA} and
\eqn{eq:AFB}.   

Let us  define the $\beta$ functions for the corresponding couplings 
  in the \drbar{} scheme:
\begin{eqnarray}
  \lefteqn{\betaDRbar_s(\asDRbar,\alpha_e,\{v_r\})
   = \mu^2 \frac{{\rm d}}{{\rm d}\mu^2} \apiDR}
   \nonumber\\
   &=& - \left[\epsilon \apiDR + 2 \frac{\asDRbar}{\ZsDRbar}
                 \left(
                 \frac{\partial \ZsDRbar}{\partial \alpha_e} \beta_e
                 + \sum_r \frac{\partial \ZsDRbar}{\partial v_r}
   \beta_{v_r} 
                 \right) \right]
   \left(1+ 2 \frac{\asDRbar}{\ZsDRbar} \frac{\partial \ZsDRbar}
   {\partial \asDRbar}\right)^{-1}    
   \nonumber\\
         &=& - \epsilon\apiDR - \sum_{i,j,k,l,m,n} \betaDRbar_{ijklmn}
   \left(\apiDR\right)^i\left(\aepi\right)^j
   \left(\frac{v_1}{\pi}\right)^k
   \left(\frac{v_2}{\pi}\right)^l 
   \left(\frac{v_3}{\pi}\right)^m 
   \left(\frac{v_4}{\pi}\right)^n
         \,,
    \label{eq::Zg_beta1}\\
   \lefteqn{\beta_e(\asDRbar,\alpha_e,\{v_r\})
                 = \mu^2 \frac{{\rm d}}{{\rm d}\mu^2} \aepi}
   \nonumber\\
   &=& - \left[\epsilon \aepi + 2 \frac{\alpha_e}{Z_e} 
                 \left(
                 \frac{\partial Z_e}{\partial \asDRbar} \betaDRbar_s
                 + \sum_r \frac{\partial Z_e}{\partial v_r} \beta_{v_r}
                 \right) \right]
   \left(1+ 2 \frac{\alpha_e}{Z_e} \frac{\partial Z_e}
   {\partial \alpha_e}\right)^{-1} 
   \nonumber\\
         &=& - \epsilon\aepi
    - \sum_{i,j,k,l,m,n} \beta^{e}_{ijklmn} 
    \left(\apiDR\right)^i
    \left(\aepi\right)^j 
    \left(\frac{v_1}{\pi}\right)^k
    \left(\frac{v_2}{\pi}\right)^l 
    \left(\frac{v_3}{\pi}\right)^m 
    \left(\frac{v_4}{\pi}\right)^n
    \,,
    \label{eq::Zg_beta2}\\
   \lefteqn{\beta_{v_r}(\asDRbar,\alpha_e,\{v_r\})
   = \mu^2 \frac{{\rm d}}{{\rm d}\mu^2} \frac{v_r}{\pi}}
   \nonumber\\
   &=&
   - \left[\epsilon \frac{v_r}{\pi}
     + 2 \frac{v_r}{Z_{\lambda_r}} 
     \left(
     \frac{\partial Z_{\lambda_r}}{\partial \asDRbar} \betaDRbar_s
     + \frac{\partial Z_{\lambda_r}}{\partial \alpha_e} \beta_e 
     + \sum_{r^\prime\not=r}
     \frac{\partial Z_{\lambda_r}}{\partial v_{r^\prime}}
     \beta_{v_{r^\prime}}  
     \right)\right]\nn
&&
   \left(1+ 2 \frac{v_r}{Z_{\lambda_r}} 
   \frac{\partial Z_{\lambda_r}}
        {\partial v_r}\right)^{-1}
                \nonumber\\
                &=& - \epsilon\frac{v_r}{\pi}
    - \sum_{i,j,k,l,m,n} \beta^{v_r}_{ijklmn} 
    \left(\apiDR\right)^i
    \left(\aepi\right)^j 
    \left(\frac{v_1}{\pi}\right)^k
    \left(\frac{v_2}{\pi}\right)^l 
    \left(\frac{v_3}{\pi}\right)^m 
    \left(\frac{v_4}{\pi}\right)^n
                \,.
    \label{eq::Zg_beta3}
\end{eqnarray}
Here and in the following we do not explicitly display the dependence
on the renormalisation scale $\mu$, i.e., $\alpha_s\equiv \alpha_s(\mu)$
etc.  Note that in the second line of Eq.\,(\ref{eq::Zg_beta1}), the
${\cal O}(\epsilon)$ terms of $\beta_e$ and $\beta_{v_r}$ contribute
to the finite part of $\betaDRbar_s$, and similarly for
Eqs.\,(\ref{eq::Zg_beta2}) and (\ref{eq::Zg_beta3}).  As we will see
below, in order to compute the four-loop term of $\betaDRbar$ one needs
$\beta_e$ to two loops and $\beta_{v_r}$ ($r=1,\cdots 4$) to one loop.

For the cases when the fermion representation allows a mass term 
we introduce the fermion mass
anomalous dimension, which is defined through
\begin{eqnarray}
  \gammaDRbar_m
  &=& \frac{\mu^2}{\mDRbar} \frac{{\rm d}}{{\rm d}\mu^2} \mDRbar
  \nonumber\\
  &=& - \pi \betaDRbar_s
  \frac{\partial \ln \ZmDRbar}{\partial \asDRbar}
  -  \pi \beta_e \frac{\partial  \ln \ZmDRbar }{\partial
  \alpha_e}
  -\pi \sum_r \beta_{v_r} \frac{\partial  \ln \ZmDRbar
  }{\partial v_r}
        \nonumber\\
        &=& - \sum_{i\cdots n} \gammaDRbar_{i\cdots n} 
  \left(\apiDR\right)^i
  \left(\aepi\right)^j 
  \left(\frac{v_1}{\pi}\right)^k
  \left(\frac{v_2}{\pi}\right)^l 
  \left(\frac{v_3}{\pi}\right)^m 
  \left(\frac{v_4}{\pi}\right)^n
  \,.
  \label{eq::Zm_gamma}
\end{eqnarray}
From this equation one can see that for the four-loop term of
$\gammaDRbar_m$, the beta functions $\beta_e$ and $\beta_{v_r}$ are
needed to three loops  and one loop, respectively,
since the dependence of $\ZmDRbar$ on $\alpha_e$ ($v_r$) starts at
one loop (three loops)~\cite{Harlander:2006rj}. The general result
for the two-loop order $\beta_e$ is known~\cite{Harlander:2006rj} and 
for the three-loop order contributions we find (generalising the 
\qcd\ results from \reference{Harlander:2006xq}, using a similar
calculational setup to the one applied
in~\cite{Harlander:2006rj,Harlander:2006xq}, which relies on the computer
programs {\tt QGRAF}~\cite{Nogueira:1991ex}, {\tt q2e}, {\tt
exp}~\cite{Seidensticker:1999bb,Harlander:1997zb} and {\tt
MINCER}~\cite{Larin:1991fz}):

\bea
\beta^e_{022000}&=& \frac{1}{256}[-6I_2(R)N_An_f-12N_AC_R+6N_AC_A
-10C_R+35C_A+15I_2(R)n_f]\nn
\beta^e_{010030}&=&-\frac{63}{1024}C_A^3\nn
\beta^e_{030100}&=&-\frac{3}{32}[C_A^2+6I_2(R)C_Rn_f-10C_AC_R+2C_AI_2(R)n_f+16
C_R^2]\nn
\beta^e_{020200}&=&-\frac{1}{256}[25N_AC_A-3I_2(R)N_An_f-30N_AC_R-5C_A-33I_2(R)n_f 
+118C_R]\nn
\beta^e_{021010}&=&\frac{1}{128}[-27I_2(R)n_f+53C_A-30C_R]C_A\nn
\beta^e_{210010}&=&-\frac{3}{1024N_AI_2(R)}[5C_A^3
I_2(R)N_A+128D_2 (RA)]\nn
\beta^e_{110101}&=&-\frac{17}{64}C_A^3\nn
\beta^e_{211000}&=&-\frac{15}{512}C_A^2\nn
\beta^e_{012100}&=&\frac{3}{64}(N_A - 1)\nn
\beta^e_{021100}&=&\frac{1}{128}[3N_AC_A+15
I_2(R)N_An_f+6N_AC_R-9C_A+3I_2(R)n_f-50C_R]\nn
\beta^e_{010003}&=&\frac{1}{24576N_A}(96D_2 (A)-7
C_A^4N_A)C_A^2\nn
\beta^e_{020110}&=&\frac{1}{128}[35C_A+27I_2(R)n_f-18C_R]C_A\nn
\beta^e_{110200}&=&\frac{1}{64}(N_A-9)C_A\nn
\beta^e_{010300}&=&-\frac{1}{256}(3N_A+2)(N_A - 1)\nn
\beta^e_{111001}&=&-\frac{3}{64}C_A^3\nn
\beta^e_{010120}&=&\frac{3}{512}C_A^2\nn
\beta^e_{010021}&=&-\frac{9}{2048N_A}(32
D_2 (A)+7C_A^4N_A)\nn
\beta^e_{110020}&=&\frac{33}{128}C_A^3
\nonumber
\eea

\bea
\beta^e_{011020}&=&-\frac{93}{512}C_A^2\nn
\beta^e_{020020}&=& \frac{1}{512}[71C_A-42C_R-81I_2(R)n_f]C_A^2\nn
\beta^e_{010012}&=&-\frac{21}{4096}C_A^5\nn
\beta^e_{010111}&=&\frac{81}{512}C_A^3\nn
\beta^e_{110011}&=&\frac{11}{128}C_A^4\nn
\beta^e_{011011}&=&-\frac{51}{512}C_A^3\nn
\beta^e_{011200}&=&\frac{3}{256}(N_A-1)(N_A-2)\nn
\beta^e_{020011}&=&\frac{1}{1536N_AI_2(R)}\{-384C_AD_2 (RA)
+768D_2 (A)I_2(R)\nn &+& N_AI_2(R)[71C_A^4
-42C_RC_A^3-81I_2(R)n_fC_A^3]\}\nn
\beta^e_{121000}&=&-\frac{1}{32}(11C_A^2-8C_AC_R+8C_R^2)\nn
\beta^e_{010210}&=&\frac{3}{256}(5N_A-2)C_A\nn
\beta^e_{110110}&=&-\frac{11}{32}C_A^2\nn
\beta^e_{011110}&=&-\frac{3}{128}(-10+N_A)C_A\nn
\beta^e_{012010}&=&-\frac{9}{64}C_A\nn
\beta^e_{111010}&=&\frac{11}{32}C_A^2\nn
\beta^e_{120010}&=&-\frac{1}{32}(14C_A+5C_R)C_A^2\nn
\beta^e_{030010}&=&-\frac{3}{64}(C_A-2C_R-9I_2(R)n_f)C_A^2\nn
\beta^e_{031000}&=&-\frac{3}{64}[-12C_AC_R+6I_2(R)C_Rn_f-7
C_AI_2(R)n_f+2C_A^2+16C_R^2]\nn
\beta^e_{210100}&=&-\frac{1}{512}(-192C_R+47C_A)C_A\nn
\beta^e_{111100}&=&-\frac{1}{32}(5N_A-1)C_A\nn
\beta^e_{010102}&=&\frac{1}{2048N_A}(96D_2 (A)-13C_A^4N_A)
\nonumber
\eea
\bea
\beta^e_{013000}&=&-\frac{1}{64}(-1+N_A)\nn
\beta^e_{110002}&=&-\frac{1}{1536N_A}[-11C_A^4N_A+192D_2 (A)]C_A\nn
\beta^e_{112000}&=&\frac{1}{64}(3N_A-5)C_A\nn
\beta^e_{011002}&=&-\frac{1}{2048N_A}(96D_2 (A)-43C_A^4N_A)\nn
\beta^e_{010201}&=&-\frac{3}{512}(5N_A-6)C_A^2\nn
\beta^e_{011101}&=&\frac{9}{256}N_AC_A^2\nn
\beta^e_{020101}&=&-\frac{1}{768N_AI_2(R)}\{N_AI_2(R)[185C_A^3
-207C_A^2I_2(R)n_f-342C_A^2C_R]\nn
&+&960D_2 (RA)\}\nn
\beta^e_{210001}&=&-\frac{1}{2048N_AI_2(R)}[5C_A^4I_2(R)N_A-384C_AD_2 (RA)
+96D_2 (A)I_2(R)]\nn
\beta^e_{012001}&=&-\frac{3}{128}C_A^2\nn
\beta^e_{021001}&=&-\frac{1}{768N_AI_2(R)}\{N_AI_2(R)[90C_A^2
C_R-143C_A^3 -63C_A^2I_2(R)n_f]\nn
&+&192D_2 (RA)\}\nn
\beta^e_{120100}&=&\frac{1}{16}(3C_A^2-8C_R^2+13C_AC_R)\nn
\beta^e_{030001}&=&-\frac{1}{128N_AI_2(R)}\{(C_A-2C_R)[-96D_2 (RA)
+C_A^3I_2(R)N_A]\nn
&-&9I_2(R)[-8D_2 (RA)+C_A^3I_2(R)N_A]n_f\}\nn
\beta^e_{020002}&=&-\frac{1}{18432N_AI_2(R)[-25C_A^4N_A+12D_2 (A)
(2+N_A)]}\nn
&&\{-3981312
D_3 (A)D_2 (RA)
+1132800C_A^6D_2 (RA)N_A
\nn
&-&829440C_A^3D_3 (A)
I_2(R)N_A+12C_A^5D_2 (A)I_2(R)(18058-71N_A)N_A\nn
&+&1775C_A^9I_2(R)N_A^2+1152C_A D_2 (A)^2I_2(R)(586+5N_A)\nn
&+&179712D_3 (RAA)[-25C_A^4N_A+12D_2 (A)(2+
N_A)]\nn
&+&9216C_A^2[540C_RD_3 (A)I_2(R)N_A
+D_2 (A) D_2 (RA)(362+N_A)]\nn
&-&75C_A^8I_2(R)N_A^2[14C_R+27I_2(R)n_f]\nn
&+&36C_A^4D_2 (A)I_2(R)N_A[2C_R(-16586+7N_A)\nn
&+&9I_2(R)(206+3N_A)n_f]
-1990656 D_3 (A) D_2 (RA) N_A
\nn
&-&3456 D_2 (A)^2I_2(R)[2C_R(602+13N_A)+9I_2(R)(2+N_A)n_f]\}\nn
\beta^e_{120001}&=&-\frac{1}{192}C_A^3(14C_A+5C_R)
+\frac{1}{8N_A I_2(R)}(5C_A-4C_R)D_2 (RA)\nonumber
\eea
\bea
\beta^e_{130000}&=&-\frac{1}{64N_AI_2(R)}\Bigl(48D_2 (RA)
+I_2(R)N_A\{11C_A^3-242
C_A^2C_R\nn
&+&640C_AC_R^2-416C_R^3
-28C_A^2I_2(R)n_f+144C_AC_RI_2(R)n_f\nn
&-&104C_R^2I_2(R)n_f
-12C_AI_2(R)^2n_f^2\nn
&+&48(C_A-2C_R)(C_A-C_R)[2C_R-C_A+I_2(R)n_f]\zeta_3\}\Bigr)\nn
\beta^e_{220000}&=&-\frac{1}{1536N_AI_2(R)}
\Bigl[3456D_2 (R)n_f+I_2(R)N_A[-
335C_A^3-642C_A^2C_R\nn
&-&2148C_AC_R^2
+3336C_R^3+3(247C_A^2+896C_AC_R-1180C_R^2)I_2(R)n_f\nn
&+&24(C_A-16C_R)I_2(R)^2n_f^2]
-384D_2 (RA)-288\Bigl(24D_2 (RA)\nn
&+&24D_2 (R)n_f
+I_2(R)N_A\{-22C_A^3
+6C_R^2[6C_R-I_2(R)n_f]\nn
&+&3C_AC_R[-32C_R+I_2(R)n_f]
+C_A^2[81C_R+2I_2(R)n_f]\}\Bigr)\zeta_3\Bigr]\nn
\beta^e_{310000}&=&-\frac{1}{13824N_AI_2(R)}\Bigl(-15552D_2 (RA)
+I_2(R)N_A\{13755C_A^
3\nn
&-&4C_A^2[13819C_R+1389I_2(R)n_f]\nn
&-&8C_R[3483C_R^2-280I_2(R)^2n_f^2+108C_RI_2(R)n_f(
-23+24\zeta_3)]\nn
&+&4C_A[12339C_R^2+120I_2(R)^2n_f^2+4C_RI_2(R)n_f(157+1296\zeta_3)]\}\Bigr)\nn
\beta^e_{040000}&=&-\frac{1}{192N_AI_2(R)[12D_2 (A)
(2+N_A)-25C_A^4N_A]}\nn
&&\Bigl[8640C_A^3D_2 (RA)I_2(R)N_An_f
-51840C_A^2C_RD_2 (RA)I_2(R)N_A
n_f\nn
&+&10368 D_2 (RA)^2(2+N_A)n_f
+100C_A^7I_2(R)N_A^2(-4+111\zeta_3)
\nn
&-&150C_A^4N_A[C_R
I_2(R)^3N_An_f^2
+2C_R^2I_2(R)^2N_An_f(23-6\zeta_3)\nn
&+&24D_2 (R)n_f(-7+2\zeta_3)+8C_R^3
I_2(R)N_A(7+9\zeta_3)\nn
&+&16D_2 (RA)(-1+12\zeta_3)]+75C_A^5I_2(R)N_A^2[7I_2(R)^2n_f^2\nn
&-&4C_R
I_2(R)n_f(-31+9\zeta_3)+8C_R^2(17+60\zeta_3)]\nn
&-&75C_A^6I_2(R)N_A^2[I_2(R)n_f(47-16\zeta_3)
+4C_R (8+117\zeta_3)]\nn
&+&12D_2 (A)\Bigl(96D_2 (RA)(2+N_A)(-1+12\zeta_3)\nn
&-&4C_A^3I_2(R)N_A(2+N_A)
(-4+111\zeta_3)\nn
&+&3C_A^2I_2(R)N_A\{4C_R(2+N_A)(8+117\zeta_3)\nn
&+&I_2(R)n_f[118+47N_A-16(2+N_A)
\zeta_3]\}\nn
&-&3C_AI_2(R)N_A\{7I_2(R)^2(2+N_A)n_f^2
+8C_R^2(2+N_A)(17+60\zeta_3)\nn
&+&4C_RI_2(R)n_f[
134+31N_A-9(2+N_A)\zeta_3]\}\nn
&+&6C_RI_2(R)^3N_A(2+N_A)n_f^2
+144D_2 (R)(2+N_A)n_f(-7+
2\zeta_3)\nn
&+&48C_R^3I_2(R)N_A(2+N_A)(7+9\zeta_3)\nn
&+&12C_R^2I_2(R)^2N_An_f[262+23N_A
-6(2+N_A)
\zeta_3]\Bigr)\Bigl].
\label{eq::betae4}
\eea
(Here and elsewhere we denote $\zeta(n)$ by $\zeta_n$.) 
We computed the four-loop \dred\ quantities from their \dreg\ counterparts
using the   
indirect method discussed in Refs.~\cite{Harlander:2006rj,Bern:2002zk}.
It is based on the following formul\ae{}:
\begin{eqnarray}
  \betaDRbar_s
  &=& \betaMSbar_s
  \frac{\partial \asDRbar}{\partial \asMSbar} +
  \beta_e \frac{\partial \asDRbar}{\partial \alpha_e} +
  \sum_r \beta_{v_r} \frac{\partial \asDRbar}{\partial v_r}
  \,,
  \nonumber\\
  \gammaDRbar_m &=&
  \gammaMSbar_m \frac{\partial \ln \mDRbar}{\partial \ln \mMSbar}  
  + \frac{\pi \betaMSbar_s}{\mDRbar}
  \frac{\partial \mDRbar}{\partial \asMSbar}
  + \frac{\pi \beta_e}{\mDRbar}
  \frac{\partial \mDRbar}{\partial \alpha_e}
  + \sum_r \frac{\pi \beta_{v_r}}{\mDRbar}  
  \frac{\partial \mDRbar}{\partial v_r}
  \,.
  \label{eq::DRED-DREG}
\end{eqnarray}
Let us briefly discuss the order in perturbation theory
up to which the individual building blocks are needed. Of course, the
\msbar{} quantities are needed to four-loop order; they can be found in
Refs.~\cite{vanRitbergen:1997va,Chetyrkin:1997dh,
Vermaseren:1997fq,Czakon:2004bu}.  The dependence of $\asDRbar$ and
$\mDRbar$ on $\alpha_e$ starts at two- and one-loop
order~\cite{Harlander:2006rj}, respectively. Thus, $\beta_e$ is needed
up to the three-loop level (cf. Eq.~(\ref{eq::DRED-DREG})).  On the other
hand, both $\asDRbar$ and $\mDRbar$ depend on $v_r$ starting from
three loops and consequently only the one-loop term of $\beta_{v_r}$
enters in Eq.~(\ref{eq::DRED-DREG}). It is given in
Eq.~(\ref{eq:genbetavs}).

For the four-loop analysis we also require 
the three-loop relations between $\asDRbar$ and $\asMSbar$ and
between $\mDRbar$ and $\mMSbar$. The two-loop results were 
presented in \reference{Harlander:2006rj}, and the 
three-loop results for the special case of \qcd\ in 
\reference{Harlander:2006xq}.  Parametrising the three-loop
terms by $\delta_\alpha^{(3)}$ and $\delta_m^{(3)}$, we have
\begin{eqnarray}
  \asDRbar &=& \asMSbar\left[1+\frac{\asMSbar}{\pi} \frac{1}{12}C_A
  +\Bigl(\frac{\asMSbar}{\pi}\right)^2
  \frac{11}{72}C_A^2
  - \frac{\asMSbar}{\pi} \aepi
  \frac{1}{8}C_RI_2(R)n_f
  + \delta_\alpha^{(3)} + \ldots \Bigr]
  \,,
  \nonumber\\
  \mDRbar &=& \mMSbar\Bigg[1 -\aepi\frac{1}{4}C_R +
  \left(\apiMS\right)^2 \frac{11}{192}C_AC_R 
-\apiMS\aepi
  \frac{1}{32}C_R(3C_A+8C_R) 
  \nonumber\\
  &&\mbox{} + \left(\aepi\right)^2 \frac{1}{32}[3C_R+I_2(R)n_f] 
      + \delta_m^{(3)} + \ldots
  \Bigg]
\eea
where the dots denote higher orders in $\asMSbar$, $\alpha_e$, and
$v_r$. We find
\bea
\pi^3\delta^{(3)}_{\alpha}
 &=& \frac{1}{96}\asMSbar \alpha_e^2I_2(R)n_f[2C_A^2-3C_AC_R+2C_R^2
-C_AI_2(R)n_f]\nn
&+&7C_RI_2(R)n_f) 
-\frac{1}{192}(\asMSbar)^2\alpha_eI_2(R)n_f(5C_A^2+60C_AC_R+6C_R^2) \nn
&+&\frac{1}{9216}\asMSbar (-
168C_A^3v_4v_2-72C_A^3v_4v_1+12v_3v_4C_A^4-48v_2v_3C_A^2\nn
&+&48
v_1v_3C_A^2-48C_Av_1v_2-48C_AN_Av_1v_2
+36C_A^3v_3^2+C_A^5v_4^2
-72C_Av_2^2\nn &-&24C_AN_Av_2^2-24C_Av_1^2)-
\frac{1}{96N_A}\asMSbar v_4^2C_AD_2 (A)
+\frac{1}{48}(\asMSbar)^2v_4D_2 (A)\nn
&+&\frac{1}{4608}(\asMSbar)^2(-6C_A^3v_3+84 
C_A^2v_2+36C_A^2v_1
-v_4C_A^4)\nn
&+&\frac{1}{10368}(\asMSbar)^3[3049C_A^3-416C_A^2I_2(R)n_f
-138C_AC_RI_2(R)n_f] 
\label{eq:deltaa}
\eea
\bea
\pi^3\delta^{(3)}_m &=& -\frac{1}{384}\alpha_e^3 C_R
[-10C_A^2+14C_AC_R+27C_R^2-7C_AI_2(R)n_f\nn
&+&39C_RI_2(R)n_f
-10I_2(R)^2n_f^2+12C_A^2\zeta_3-36C_AC_R\zeta_3+24C_R^2\zeta_3]\nn
&-&\alpha_e^2 C_R\Bigl(\frac{1}{192}[6C_Rv_1
+12C_Rv_2-2C_Av_2-C_Av_1]\nn
&+&\frac{1}{16I_2(R)N_A}D_2 (RA)v_4
+\frac{1}{384}\asMSbar[47C_A^2
+10C_R^2\nn
&-&3I_2(R)C_An_f
-19I_2(R)C_Rn_f-165C_AC_R+144C_R^2\zeta_3\nn
&-&48
I_2(R)C_An_f\zeta_3
+48I_2(R)C_Rn_f\zeta_3
+72C_A^2\zeta_3
-216C_AC_R\zeta_3]\Bigr)\nn
&+&\alpha_e C_R
\Bigl(\frac{1}{12288}[200v_2^2+88N_Av_2^2+56v_1^2
+16N_Av_1^2\nn
&+&112N_Av_2v_1-C_A^4v_4^2-12C_A^3v_3v_4+176v_2v_1
+48v_3C_Av_2\nn
&-&36C_A^2v_3^2+488C_A^2v_4v_2+232C_A^2v_4
v_1-48C_Av_1v_3]\nn
&+&\frac{1}{3072}(\asMSbar)^2[2880C_R^2\zeta_3-168C_AI_2(R)n_f
-1544C_AC_R-52C_R^2\nn
&-&128I_2(R)C_Rn_f
+1440C_A^2\zeta_3-4320C_AC_R\zeta_3
-79C_A^2]\Bigr)\nn
&+&\frac{1}{20736}(\asMSbar)^3 C_RC_A[4354C_A+135C_R+304I_2(R)
n_f]\nn
&+&\frac{3}{128N_A}D_2 (A)v_4^2.
\label{eq:deltam}
\eea

\subsection{The $\beta$ function and anomalous dimension}
Inserting Eqs.~(\ref{eq:deltaa}) and (\ref{eq:deltam}) into 
Eq.~(\ref{eq::DRED-DREG}), we obtain
\bea
\betaDRbar_{500000}&=&b_3 - \frac{1}{165888N_A}\{2592D_2 (A) 
\nn&+&
     C_AN_A[27648b_2 - C_A(1152b_1 + 85280b_0C_A + 27C_A^2) \nn&+&
       64b_0(208C_A + 69C_R)I_2(R)n_f]\} \nn 
\betaDRbar_{203000}&=&\frac{1}{192}C_A\nn
\betaDRbar_{302000}&=&-\frac{1}{64}C_A^2\nn
\betaDRbar_{311000}&=&-\frac{1}{128}n_fI_2(R)C_A^2\nn
\betaDRbar_{300002}&=&-\frac{1}{663552N_A}[19872C_A^2D_2 (A)
-227C_A^6N_A+27648D_3 (A)]\nn
\betaDRbar_{400010}&=&-\frac{1}{3072}(4b_0+9C_A)C_A^3\nn
\betaDRbar_{200030}&=&-\frac{11}{3072}C_A^4\nn
\betaDRbar_{310001}&=&-\frac{1}{4608N_A}[96D_2 (A)-C_A^4N_A]I_2(R)n_f\nn
\betaDRbar_{220100}&=&-\frac{1}{48}(C_A+3C_R)n_fI_2(R)C_R\nn
\betaDRbar_{301100}&=&-\frac{1}{256}(5N_A+12)C_A^2\nn
\betaDRbar_{210011}&=&-\frac{1}{1536}(4C_A+3C_R)n_fC_A^3I_2(R)\nn
\betaDRbar_{400001}&=&\frac{1}{18432N_A}(4b_0+9C_A)[96D_2 (A)-C_A^4N_A]\nn
\betaDRbar_{300011}&=&\frac{1}{55296N_A}[384D_2 (A)+227C_A^4N_A]C_A\nn
\betaDRbar_{400100}&=&\frac{7}{1536}(4b_0+9C_A)C_A^2\nn
\betaDRbar_{401000}&=&\frac{1}{512}(4b_0+9C_A)C_A^2\nn
\betaDRbar_{200120}&=&-\frac{3}{512}C_A^3\nn
\betaDRbar_{201020}&=&-\frac{7}{512}C_A^3\nn
\betaDRbar_{310100}&=&-\frac{7}{384}n_fI_2(R)C_A^2\nonumber
\eea
\bea
\betaDRbar_{220001}&=&-\frac{1}{1152N_A[-25C_A^4N_A
+ 12D_2 (A)(2 + N_A)]}\nn
&&
\{C_A^3(C_A + 3C_R)(25C_A^4 - 12D_2 (A))
         I_2(R)N_A^2\nn&-&
        24(C_A + 3C_R)[25C_A^4D_2 (RA)
- 36 D_2 (A) D_2 (RA)\nn 
&+&
          C_A^3D_2 (A)I_2(R)]N_A\}  \nn
\betaDRbar_{202100}&=&\frac{1}{64}(N_A+1)C_A\nn
\betaDRbar_{410000}&=&-\frac{1}{1536}I_2(R)n_f\{8b_0(5C_A^2 + 56C_AC_R
+ 6C_R^2) \nn&+&
      C_R[-192b_1 + 39C_A^2 + 892C_AC_R + 108C_R^2 + 24C_AI_2(R)n_f \nn&-&
        80C_RI_2(R)n_f]\} \nn
\betaDRbar_{301001}&=&-\frac{1}{3072}[96D_2 (A)+89C_A^4N_A]\nn
\betaDRbar_{220010}&=&\frac{1}{192}(C_A+3C_R)n_fI_2(R)C_A^2\nn
\betaDRbar_{300020}&=&\frac{1}{18432N_A}[96D_2 (A)+227C_A^4N_A]\nn
\betaDRbar_{210110}&=&\frac{1}{384}(4C_A+3C_R)n_fI_2(R)C_A\nn
\betaDRbar_{320000}&=&-\frac{1}{1152N_A}n_f\Bigl(-24D_2 (RA) \nn&+&
      I_2(R)N_A\{16C_A^3 + 6C_AC_R[25C_R - 22I_2(R)n_f] \nn&+&
        3C_A^2[4C_R - 5I_2(R)n_f] - 72C_R^2[7C_R + 5I_2(R)n_f]\}\Bigr)\nn
\betaDRbar_{230000}&=&-\frac{1}{192}I_2(R)n_f[-4C_A^3 
+ 23C_A^2C_R - 46C_AC_R^2 + 32C_R^3 \nn&+&  
      (6C_A^2 - 33C_AC_R + 50C_R^2)I_2(R)n_f
- 2(C_A - 7C_R)I_2(R)^2n_f^2] \nn
\betaDRbar_{300200}&=&-\frac{1}{1536}(19N_A+82)C_A^2\nn
\betaDRbar_{200111}&=&-\frac{1}{512}C_A^4\nn
\betaDRbar_{201011}&=&-\frac{9}{512}C_A^4\nn
\betaDRbar_{300110}&=&-\frac{23}{1536}C_A^3\nn
\betaDRbar_{301010}&=&\frac{11}{512}C_A^3\nn
\betaDRbar_{211100}&=&\frac{1}{384}(N_A+1)(4C_A+3C_R)I_2(R)n_f\nn
\betaDRbar_{300101}&=&-\frac{1}{9216N_A}[480D_2 (A)+703C_A^4N_A]\nonumber
\eea
\bea
\betaDRbar_{210020}&=&-\frac{1}{512}(4C_A+3C_R)n_fI_2(R)C_A^2\nn
\betaDRbar_{310010}&=&\frac{1}{768}I_2(R)C_A^3n_f\nn
\betaDRbar_{200003}&=&-\frac{1}{663552N_A}[864C_A^2D_2 (A)+11C_A^6N_A
-27648D_3 (A)]C_A\nn
\betaDRbar_{200102}&=&\frac{1}{6144N_A}[480D_2 (A)+199C_A^4N_A]C_A\nn
\betaDRbar_{201002}&=&\frac{1}{2048N_A}[96D_2 (A)+17C_A^4N_A]
C_A\nn
\betaDRbar_{202001}&=&\frac{3}{128}C_A^3\nn
\betaDRbar_{200300}&=&\frac{1}{768}(N_A^2+13N_A+18)C_A\nn
\betaDRbar_{201101}&=&\frac{3}{256}(8+N_A)C_A^3\nn
\betaDRbar_{200021}&=&-\frac{1}{6144N_A}[96D_2 (A)+11C_A^4N_A]C_A\nn
\betaDRbar_{202010}&=&-\frac{1}{64}C_A^2\nn
\betaDRbar_{210002}&=&\frac{1}{18432N_A}(4C_A+3C_R)[96D_2 (A)-C_A^4N_A]I_2(R)n_f\nn
\betaDRbar_{212000}&=&\frac{1}{768}(4C_A+3C_R)I_2(R)n_f\nn
\betaDRbar_{211010}&=&-\frac{1}{384}(4C_A+3C_R)n_fI_2(R)C_A\nn
\betaDRbar_{201200}&=&\frac{1}{256}(N_A^2+5N_A+10)C_A\nn
\betaDRbar_{200210}&=&\frac{1}{256}(N_A-2)C_A^2\nn
\betaDRbar_{201110}&=&-\frac{1}{128}(N_A-2)C_A^2\nn
\betaDRbar_{221000}&=&\frac{1}{192}(C_A+3C_R)(C_A-2C_R)I_2(R)n_f\nn
\betaDRbar_{200201}&=&\frac{1}{512}(7N_A+46)C_A^3\nn
\betaDRbar_{211001}&=&\frac{1}{256}(4C_A+3C_R)n_fI_2(R)C_A^2\nn
\betaDRbar_{200012}&=&-\frac{1}{36864N_A}[384D_2 (A)+11C_A^4N_A]C_A^2\nn
\betaDRbar_{210200}&=&\frac{1}{768}(N_A+3)(4C_A+3C_R)I_2(R)n_f\nn
\betaDRbar_{210101}&=&\frac{7}{768}(4C_A+3C_R)n_fI_2(R)C_A^2
\eea

where
\bea
b_0&=&\frac{1}{4}\Bigl(\frac{11}{3}C_A-\frac{4}{3}I_2(R)n_f\Bigr)\nn
b_1&=&\frac{1}{16}\Bigl(\frac{34}{3}C_A^2-4C_RI_2(R)n_f
-\frac{20}{3}C_AI_2(R)n_f\Bigr)\nn
b_2&=&\frac{1}{64}\Bigl(\frac{2857}{54}C_A^3
+2C_R^2I_2(R)n_f-\frac{205}{9}C_RC_AI_2(R)n_f\nn
        &-&\frac{1415}{27}C_A^2I_2(R)n_f+\frac{44}{9}C_RI_2(R)^2n_f^2
+\frac{158}{27}C_AI_2(R)^2n_f^2\Bigr)\nn
b_3&=&\frac{1}{256}\Bigl[
\Bigl(\frac{150653}{486}-\frac{44}{9}\zeta_3\Bigr)C_A^4
+C_A^3I_2(R)n_f\Bigl(-\frac{39143}{81}+\frac{136}{3}\zeta_3\Bigr)\nn
&+&C_A^2C_RI_2(R)n_f\Bigl(\frac{7073}{243}-\frac{656}{9}\zeta_3\Bigr)
+C_AC_R^2I_2(R)n_f\Bigl(-\frac{4204}{27}+\frac{352}{9}\zeta_3\Bigr)\nn
&+&46C_R^3I_2(R)n_f+C_A^2I_2(R)^2n_f^2\Bigl(\frac{7930}{81}+\frac{224}{9}\zeta_3\Bigr)
\nn
&+&C_R^2I_2(R)^2n_f^2\Bigl(\frac{1352}{27}-\frac{704}{9}\zeta_3\Bigr)
+C_AC_RI_2(R)^2n_f^2\Bigl(\frac{17152}{243}+\frac{448}{9}\zeta_3\Bigr)\nn
&+&\frac{424}{243}C_A I_2(R)^3n_f^3 + \frac{1232}{243}C_RI_2(R)^3n_f^3\nn
&+&\frac{D_2 (A)}{N_A}\Bigl(-\frac{80}{9}
+\frac{704}{3}\zeta_3\Bigr)+
\frac{n_fD_2 (RA)}{N_A}\Bigl(\frac{512}{9}
-\frac{1664}{3}\zeta_3\Bigr)\nn
&+&\frac{n_f^2D_2 (R)}{N_A}\Bigl(-\frac{704}{9}
+\frac{512}{3}\zeta_3\Bigr)\Bigr]
\eea
are the one, two, three and four-loop gauge $\beta$-function 
coefficients 
calculated in DREG. For the fermion mass anomalous dimension, we find 
\bea
\gammaDRbar_{400000}&=& \gamma_3 + 
\frac{91}{768}C_R^2C_A^2-\frac{129}{512}C_AC_R^3
-\frac{3}{16}I_2(R)C_AC_R^2n_f\zeta_3\nn
&+&\frac{89}{576}I_2(R)C_AC_R^2n_f
+\frac{29}{5184}I_2(R)^2C_AC_Rn_f^2
+\frac{3}{16}I_2(R)C_A^2C_Rn_f\zeta_3\nn
 &-&
\frac{53}{1296}C_RC_A^2I_2(R)n_f
-\frac{19003}{82944}C_A^3C_R\nn
\gammaDRbar_{110002}&=&\frac{1}{24576N_A}[2784C_AD_2 (A)+1632C_R
D_2 (A)-53C_A^5N_A\nn
&-&11C_RC_A^4N_A]C_R\nn
\gammaDRbar_{110011}&=&-\frac{1}{2048}(53C_A+11C_R)C_A^3C_R\nn
\gammaDRbar_{120010}&=&\frac{1}{256}(31C_A+18C_R)C_A^2C_R\nn
\gammaDRbar_{210010}&=&\frac{3}{2048N_AI_2(R)}[3C_A^3I_2(R)N_A
+64D_2 (RA)]C_R\nn
\gammaDRbar_{111001}&=&\frac{1}{1024}(53C_R+79C_A)C_A^2C_R\nn
\gammaDRbar_{020110}&=&-\frac{1}{256}[17C_A+17I_2(R)n_f-6C_R]C_AC_R\nn
\gammaDRbar_{020011}&=&-\frac{1}{3072N_AI_2(R)}\{384D_2 (A)I_2(R)
-144C_AD_2 (RA)\nn
&+&N_AI_2(R)[-51C_A^3I_2(R)n_f-30C_A^3C_R+37C_A^4]\}C_R\nn
\gammaDRbar_{310000}&=&-\frac{1}{165888I_2(R)N_A}
C_R\Bigl(46656D_2 (RA) \nn&+& I_2(R)N_A\{-26505C_A^3 +
          C_A^2[355107C_R + 23544I_2(R)n_f] \nn&+& 2C_A[11916C_R^2 -
            65508C_RI_2(R)n_f - 3744I_2(R)^2n_f^2] \nn&+&
          12C_R[7965C_R^2 - 7212C_RI_2(R)n_f - 224I_2(R)^2n_f^2] \nn&+&
          2592(C_A - C_R)[5(-11C_A - 9C_R)(C_A - 2C_R)\nn
&+& 4(5C_A - 16C_R)
             I_2(R)n_f]\zeta_3\}\Bigr) \nn
\gammaDRbar_{110020}&=&-\frac{3}{2048}(53C_A+11C_R)C_A^2C_R\nn
\gammaDRbar_{211000}&=&-\frac{7}{1024}C_A^2C_R\nn
\gammaDRbar_{020020}&=&-\frac{1}{1024N_AI_2(R)}\{N_AI_2(R)[-51C_A^2I_2(R)n_f
-30C_A^2C_R\nn
&+&37C_A^3]+16D_2 (RA)\}C_R\nn
\gammaDRbar_{020101}&=&\frac{1}{1536N_AI_2(R)}\{N_AI_2(R)[103C_A^3
-154C_A^2C_R\nn
&-&277C_A^2I_2(R)n_f]+720D_2 (RA)\}C_R
\nonumber
\eea
\bea
\gammaDRbar_{110200}&=&\frac{1}{1024}(19C_RN_A+21C_AN_A+95C_A+49C_R)C_R\nn
\gammaDRbar_{021100}&=&-\frac{1}{768}[43I_2(R)N_An_f+10C_RN_A
+3C_AN_A+35I_2(R)n_f-142C_R\nn
&-&5C_A]C_R\nn
\gammaDRbar_{021010}&=&-\frac{1}{768}[-51I_2(R)n_f-42C_R+79C_A]C_AC_R\nn
\gammaDRbar_{110101}&=&\frac{1}{1024}(117C_R+211C_A)C_A^2C_R\nn
\gammaDRbar_{120100}&=&-\frac{1}{32}(C_A^2+6C_R^2+10C_AC_R)C_R\nn
\gammaDRbar_{200011}&=&-\frac{1}{1024}C_A^4C_R\nn
\gammaDRbar_{200101}&=&\frac{7}{512}C_A^3C_R\nn
\gammaDRbar_{010003}&=&\frac{1}{442368N_A}[-41472D_3 (A)+480C_A^2
D_2 (A)+37C_A^6N_A]C_R\nn
\gammaDRbar_{021001}&=&\frac{1}{1536N_AI_2(R)}\{N_AI_2(R)[-63C_A^3
-113C_A^2I_2(R)n_f+34C_A^2C_R]\nn
&+&240D_2 (RA)\}C_R\nn
\gammaDRbar_{120001}&=&\frac{1}{1536}C_R[C_A^3(31C_A + 18C_R)
- \frac{24}{I_2(R)N_A}(19C_A + 12C_R)D_2 (RA)
           ] \nn
\gammaDRbar_{010120}&=&\frac{29}{3072}C_A^2C_R\nn
\gammaDRbar_{210100}&=&-\frac{1}{1024}(96C_R+7C_A)C_AC_R\nn
\gammaDRbar_{010111}&=&-\frac{97}{3072}C_A^3C_R\nn
\gammaDRbar_{300010}&=&\frac{1}{1024}C_A^3C_R\nn
\gammaDRbar_{300100}&=&-\frac{7}{512}C_A^2C_R\nn
\gammaDRbar_{301000}&=&-\frac{3}{512}C_A^2C_R\nn
\gammaDRbar_{011002}&=&-\frac{1}{36864N_A}[3360D_2 (A)+1033C_A^4N_A]C_R\nn
\gammaDRbar_{010012}&=&\frac{1}{24576N_A}[384D_2 (A)+37C_A^4N_A]C_RC_A\nn
\gammaDRbar_{010300}&=&-\frac{1}{1536}(N_A^2+57N_A+86)C_R\nn
\gammaDRbar_{012001}&=&-\frac{1}{1536}(3N_A+74)C_A^2C_R\nn
\gammaDRbar_{011011}&=&\frac{49}{1024}C_A^3C_R\nonumber
\eea
\bea
\gammaDRbar_{111010}&=&-\frac{1}{512}(53C_A+11C_R)C_AC_R\nn
\gammaDRbar_{200020}&=&-\frac{3}{1024}C_A^3C_R\nn
\gammaDRbar_{201010}&=&-\frac{1}{256}C_A^2C_R\nn
\gammaDRbar_{200002}&=&-\frac{1}{12288N_A}[-96D_2 (A)
+C_A^4N_A]C_RC_AN_A\nn
\gammaDRbar_{202000}&=&\frac{1}{512}C_AC_R\nn
\gammaDRbar_{201001}&=&\frac{3}{512}C_A^3C_R\nn
\gammaDRbar_{220000}&=&\frac{1}{18432I_2(R)N_A}
C_R\{-288D_2 (RA)(1 + 72\zeta_3)\nn
&+&2C_A^3I_2(R)N_A(-1295 + 10080\zeta_3) \nn&+&
        4[3C_R^3I_2(R)N_A(1544 - 5760\zeta_3) +
2592D_2 (R)n_f(1 - 2\zeta_3) \nn&+&
          64C_RI_2(R)^3N_An_f^2(-10 + 3\zeta_3) - 2C_R^2I_2(R)^2N_An_f
           (571 + 1008\zeta_3)] \nn&+& C_A^2I_2(R)N_A[2354I_2(R)n_f
- 2784I_2(R)n_f\zeta_3 -
          12C_R(721 + 7704\zeta_3)]\nn
&+& 2C_AI_2(R)N_A[24I_2(R)^2n_f^2(17 - 16\zeta_3) \nn
&+&
          C_RI_2(R)n_f(7444 + 5856\zeta_3) + C_R^2(9428 + 71136\zeta_3)]\}\nn
\gammaDRbar_{011020}&=&\frac{181}{3072}C_A^2C_R\nn
\gammaDRbar_{010021}&=&\frac{1}{4096N_A}[224D_2 (A)+37C_A^4N_A]C_R\nn
\gammaDRbar_{200200}&=&\frac{1}{512}(N_A+3)C_AC_R\nn
\gammaDRbar_{201100}&=&\frac{1}{256}(N_A+1)C_AC_R\nn
\gammaDRbar_{300001}&=&\frac{1}{6144N_A}[-96D_2 (A)
+C_A^4N_A]C_R\nn
\gammaDRbar_{200110}&=&\frac{1}{256}C_A^2C_R\nn
\gammaDRbar_{111100}&=&\frac{1}{512}(37C_AN_A+15C_RN_A+19C_R+21C_A)C_R\nn
\gammaDRbar_{010030}&=&\frac{37}{2048}C_A^3C_R\nn
\gammaDRbar_{020200}&=&\frac{1}{1536}[-35I_2(R)N_An_f+39C_AN_A-38C_RN_A+302C_R
\nn
&-&121I_2(R)n_f
-23C_A]C_R\nn
\gammaDRbar_{121000}&=&\frac{1}{256}(-24C_R^2-22C_AC_R+27C_A^2)C_R\nonumber
\eea
\bea
 \gammaDRbar_{040000}&=&-\frac{1}{1536I_2(R)N_A[-25C_A^4N_A +
      12D_2 (A)(2 + N_A)]}\nn &&C_R\Bigl[
-16320C_A^3D_2 (RA)I_2(R)N_An_f \nn&+&
      97920C_A^2C_RD_2 (RA)I_2(R)N_An_f
- 19584 D_2 (RA)^2(2 + N_A)n_f \nn&+&
      100C_A^5I_2(R)N_A^2[-288C_RI_2(R)n_f - 15I_2(R)^2n_f^2 \nn&+&
        2C_R^2(-126 - 432\zeta_3)] 
- 100C_A^7I_2(R)N_A^2(-23 + 240\zeta_3)      
\nn&+&
      100C_A^6I_2(R)N_A^2\{15C_R + 75I_2(R)n_f - 6[-132C_R + I_2(R)n_f]
\zeta_3\} \nn&+&
      25C_A^4N_A\{1344C_R^2I_2(R)^2N_An_f + 180C_RI_2(R)^3N_An_f^2 \nn&-& 
        12n_f[5I_2(R)^4N_An_f^2 + 24D_2 (R)(7 - 2\zeta_3)] \nn&+&
        192D_2 (RA)(-1 + 12\zeta_3) + 3C_R^3I_2(R)N_A(368 +
384\zeta_3)\} \nn&-&
      12D_2 (A)\Bigl(
-4032D_2 (R)n_f + N_A\{1104C_R^3I_2(R)(2 + N_A) \nn&-&
          6[336D_2 (R) - 16C_R^2I_2(R)^2(79 + 14N_A)]n_f \nn&+&
          180C_RI_2(R)^3(2 + N_A)n_f^2 - 60I_2(R)^4(2 + N_A)n_f^3\}\nn&+&
        4C_AI_2(R)N_A[-24C_RI_2(R)(41 + 12N_A)n_f - 15I_2(R)^2(2 + N_A)n_f^2 
\nn&+&
          2C_R^2(2 + N_A)(-126 - 432\zeta_3)] + 4C_A^3I_2(R)N_A(2 + N_A)
         (23 - 240\zeta_3) \nn&-& 96(2 + N_A)[-12C_R^3I_2(R)N_A
- 6D_2 (R)n_f]\zeta_3 \nn&+&
        192D_2 (RA)(2 + N_A)(-1 + 12\zeta_3) \nn
&+& 4C_A^2I_2(R)N_A
         \{C_R(2 + N_A)(15 + 792\zeta_3) \nn&+& I_2(R)n_f[184 + 75N_A -
            6(2 + N_A)\zeta_3]\}\Bigr)\Bigr]\nn
\gammaDRbar_{010201}&=&-\frac{1}{1024}(23N_A+218)C_A^2C_R\nn
\gammaDRbar_{010210}&=&-\frac{3}{512}(3N_A-2)C_AC_R\nn
\gammaDRbar_{112000}&=&-\frac{1}{1024}(8C_AN_A-2C_RN_A-37C_A-15C_R)C_R\nn
\gammaDRbar_{110110}&=&\frac{1}{512}(53C_A+11C_R)C_AC_R\nn
\gammaDRbar_{130000}&=&-\frac{1}{3072I_2(R)N_A}C_R\Bigl(-576D_2 (RA) +
        I_2(R)N_A\{352C_A^3 + 616C_A^2C_R\nn 
&-&6040C_AC_R^2 + 6992C_R^3 -
          100C_A^2I_2(R)n_f - 1388C_AC_RI_2(R)n_f\nn 
&+& 3920C_R^2I_2(R)n_f +
          168C_AI_2(R)^2n_f^2 -88C_RI_2(R)^2n_f^2 \nn&+& 48(C_A - C_R)
           [3(8C_A - 13C_R)(C_A - 2C_R) \nn
&-& 2(16C_A - 32C_R)I_2(R)n_f +
            8I_2(R)^2n_f^2]\zeta_3\}\Bigr) \nn
\gammaDRbar_{022000}&=&-\frac{1}{1536}[8C_AN_A-16C_RN_A-4I_2(R)N_An_f
+43I_2(R)n_f\nn
&-&50C_R+57C_A]C_R\nonumber
\eea
\bea
\gammaDRbar_{020002}&=&\frac{1}{36864I_2(R)N_A[-25C_A^4N_A
+ 12D_2 (A)(2 + N_A)]}\nn
&&      C_R\Bigl(-1880064 D_3 (A) D_2 (RA)
+ 536400C_A^6D_2 (RA)
         N_A \nn&-& 391680C_A^3D_3 (A)I_2(R)N_A
+ 4C_A^5D_2 (A)I_2(R)
         (22778 - 111N_A)N_A \nn&+& 925C_A^9I_2(R)N_A^2
+ 1152C_A D_2 (A)^2I_2(R)
         (286 + 7N_A) \nn&+&
 89856D_3 (RAA)[-25C_A^4N_A +
          12D_2 (A)(2 + N_A)] \nn
&+& 192C_A^2[12240C_RD_3 (A)I_2(R)N_A+
          D_2 (A) D_2 (RA) (8198 + 19N_A)] \nn
&-& 75C_A^8I_2(R)N_A^2
         [10C_R +17I_2(R)n_f] \nn&+& 12C_A^4D_2 (A)I_2(R)N_A
         [-C_R(43540 - 30N_A) \nn
&+& 3I_2(R)(2634 + 17N_A)n_f] +
        3456\{-272 D_3 (A) D_2 (RA) N_A
\nn&+& D_2 (A)^2I_2(R)
           [-2C_R(290 + 9N_A) - 13I_2(R)(2 + N_A)n_f]\}\Bigr)\nn
\gammaDRbar_{030010}&=&\frac{1}{256}[3C_A-10C_R-30I_2(R)n_f]C_A^2C_R\nn
\gammaDRbar_{011110}&=&\frac{1}{768}(11N_A-50)C_AC_R\nn
\gammaDRbar_{012010}&=&\frac{1}{768}(N_A+38)C_AC_R\nn
\gammaDRbar_{030001}&=&-\frac{1}{1536N_AI_2(R)}\{384C_AD_2 (RA)
-864C_RD_2 (RA)\nn
&+&N_AI_2(R)[10C_A^3C_R-3C_A^4+30C_A^3
I_2(R)n_f]\nn
&-&528D_2 (RA)I_2(R)n_f\}C_R\nn
\gammaDRbar_{010102}&=&-\frac{1}{36864N_A}[7008D_2 (A)+2537C_A^4N_A]C_R\nn
\gammaDRbar_{030100}&=&\frac{1}{384}[13C_A^2
+8C_AI_2(R)n_f+216C_R^2+132C_RI_2(R)n_f\nn
&-&122C_AC_R]C_R\nn
\gammaDRbar_{031000}&=&\frac{1}{384}[11C_A^2+108C_R^2-41C_AI_2(R)n_f
-76C_AC_R+66C_RI_2(R)n_f]C_R\nn
\gammaDRbar_{210001}&=&-\frac{3}{4096N_AI_2(R)}(64C_AD_2 (RA)+32D_2 (A)I_2(R)-C_A^4I_2(R)N_A)C_R\nn
\gammaDRbar_{011101}&=&-\frac{1}{1536}(53N_A+320)C_A^2C_R\nn
\gammaDRbar_{011200}&=&-\frac{1}{512}(5N_A^2+21N_A+46)C_R\nn
\gammaDRbar_{012100}&=&-\frac{1}{768}(N_A^2+31N_A+22)C_R\nn
\gammaDRbar_{013000}&=&\frac{1}{768}(N_A-10)C_R\nn
\eea
where
\bea
\gamma_3  &=&\frac{1}{256}\Bigl[
C_R^4\Bigl(-\frac{1261}{8}-336\zeta_3\Bigr)
+C_R^3C_A\Bigl(\frac{15349}{12}+316\zeta_3\Bigr)\nn
&+&C_R^2C_A^2\Bigl(-\frac{34045}{36}-152\zeta_3+440\zeta_5\Bigr)
+C_RC_A^3\Bigl(\frac{70055}{72}+\frac{1418}{9}\zeta_3-440\zeta_5\Bigr)\nn&+&
C_R^3I_2(R)n_f\Bigl(-\frac{280}{3}+552\zeta_3-480\zeta_5\Bigr)\nn
&+&C_R^2C_AI_2(R)n_f\Bigl(-\frac{8819}{27}+368\zeta_3-264\zeta_4+80\zeta_5\Bigr)\nn
&+&C_RC_A^2I_2(R)n_f\Bigl(-\frac{65459}{162}-\frac{2684}{3}\zeta_3
+264\zeta_4+400\zeta_5\Bigr)\nn
&+&C_R^2I_2(R)^2n_f^2\Bigl(\frac{304}{27}-160\zeta_3+96\zeta_4\Bigr)\nn
&+&C_RI_2(R)^3n_f^3\Bigl(-\frac{664}{81}+\frac{128}{9}\zeta_3\Bigr)
+\frac{D_2 (RA)}{d_R}\Bigl(-32+240\zeta_3\Bigr)\nn
&+&C_RC_AI_2(R)^2n_f^2\Bigl(\frac{1342}{81}+160\zeta_3-96\zeta_4\Bigr)
+\frac{n_fD_2 (R)}{d_R}\Bigl(64-480\zeta_3\Bigr)\Bigr].
\eea
\section{\label{sec::moresusy}The four-loop supersymmetric case}

Our conventions are such that substituting in the above
equations  the results
of Tables~\ref{gaminvsa}-\ref{gaminvsc} corresponds to a gauge theory with
$n_f$ sets of Dirac fermions transforming according to the
fundamental representation, or $n_f$ sets of fundamental
two component fermions with $n_f$ sets of anti-fundamental
two component fermions.

To extract the \sic\ case we must make the replacements
\bea
C_R &\to& C_A\nn
I_2 (R) &\to& C_A\nn   
n_f &\to& \frak{1}{2}\nn
D_2 (R) &\to& D_2 (A)\nn
D_2 (RA) &\to& D_2 (A)\nn 
D_3 (RAA) &\to& D_3 (A)\nn
\alpha_e &=& v_3 = \alpha_s\nn
v_1 &=& v_2 = v_4 = 0.
\label{susycase}
\eea
With these substitutions we can compare
our results for $\beta_s$ with
the four-loop results for the gauge $\beta$-function
$\beta_s^{\rm SYM}$ of \sqcd\ which was presented in~\reference{Jack:1998uj}:
\begin{equation}
\begin{split}
\beta_s^{\rm SYM} = -\left(\api\right)^2\,\left[
  \frac{3}{4}C_A
  + \frac{3}{8}C_A^2\api
  + \frac{21}{64}C_A^3\left(\api\right)^2
  + \frac{51}{128}C_A^4\left(\api\right)^3
\right] + {\cal O}(\alpha_s^6)\,.
\end{split}
\label{eq::betag4}
\end{equation}
 We indeed find that using \eqn{susycase}\ in \eqn{eq::Zg_beta1}\
precisely reproduces \eqn{eq::betag4}.   

We have also checked that in the same \sic\ limit, 
\eqn{eq::betae4}\ reproduces the three-loop \sqcd\ $\beta$-function.

Turning now to the case of softly-broken \sy, 
there exists an exact result for
$\gamma_m$~\cite{Jack:1997pa}:
\begin{equation}
\gamma_m^{\rm SYM} = \pi \alpha_s\frac{\rm d}{\rm d\alpha_s}
\left[\frac{\beta_s^{\rm SYM}}{\alpha_s}\right],
\label{eq::exactgamma}
\end{equation}
whence it follows that
\begin{equation}
\begin{split}
\gamma_m^{\rm SYM} = -\left(\api\right)\,\left[
  \frac{3}{4}C_A
  + \frac{3}{4}C_A^2\api
  + \frac{63}{64}C_A^3\left(\api\right)^2
  + \frac{51}{32}C_A^4\left(\api\right)^3
\right] + {\cal O}(\alpha_s^5)\,.
\end{split}
 \label{eq::gamma4}
\end{equation}
Using \eqn{susycase}\ in \eqn{eq::Zm_gamma}\
precisely reproduces \eqn{eq::gamma4} in similar fashion.   

The invariant $D_3 (A)$ does not feature in either  calculation, and the
dependence on $D_2 (A)$, $N_A$, $\zeta_3$,  $\zeta_4$ and $\zeta_5$ all 
 cancel, although they appear in individual terms. It is tempting  to
speculate that this absence of higher order invariants and
transcendental  numbers (other than $\pi$) is related to the existence
of the \nsvz{} scheme, in which  the gauge $\beta$-function
for any simple gauge group  is given (in the supersymmetric case without
matter fields)  by the expression~\cite{Jones:1983ip}, \cite{Novikov:1983ee} 
\be \beta_s^{\rm
NSVZ} = -\frac{3}{4}C_A\left(\api\right)^2
\left(1-\frac{C_A\alpha_s}{2\pi}\right)^{-1} 
\ee 
which is manifestly
free of them to all orders. It is natural  to conjecture that the same
property holds in the \dred{} scheme.   For discussion of the
relationship between $\beta_s^{\rm NSVZ}$ and  $\betaDRbar_s$
see~\reference{cjn}.

\section{\label{sec::concl}Discussion}

In this paper we have applied \dred{} to gauge theories
with gauge groups $SU(N)$, $SO(N)$ and $Sp(N)$, and calculated
both the gauge $\beta$-function and the mass anomalous dimension
to the four-loop level. These calculations required careful
treatment of the evanescent Yukawa and quartic couplings of the
\epscalar{}.  In the \sic{} limit we explicitly verified that
the $\beta$-function for the evanescent Yukawa coupling reproduces the
gauge $\beta$-function through three loops.

The results for $\betaDRbar_s$ and $\gammaDRbar_m$ 
in the special case of \qcd{} as described in 
\reference{Harlander:2006rj} and \cite{Harlander:2006xq} are  
easily obtained from the results of this paper by specialising to the 
$SU(N)$ case, and setting $N=3$  and 
$v_4 = 0$, with the fermions in the fundamental representation. 

Predictions based on theories with low energy \sy\ require careful 
consideration of the transition between the \msbar\ and \drbar\ 
renormalisation schemes. If, for example, the decoupling of \sic\ particles is 
carried out in several steps (as in split \sy, 
for example~\cite{Arkani-Hamed:2004fb}) then it is essential to 
take into account 
the evanescent couplings (for a recent discussion and treatment of the 
running of $\alpha_s$ and $m_b$ in the \mssm, see 
\reference{hms}.

\appendix

\section{\label{appA}Group Theory}

We consider a gauge group ${\cal G}$ with generators $R^a$ 
satisfying\footnote{Useful sources for some of the material in this 
section have included Refs.~\cite{predrag,vanRitbergen:1997va,RSV}.}
\be
\left[R^a, R^b \right] = if^{abc}R^c.
\ee
We work throughout with a  fermion representation consisting of 
$n_f$ sets of Dirac fermions or $2n_f$ sets of two-component fermions, in 
irreducible representations with identical Casimirs, using $R^a$ to denote 
the generators in one such representation. Thus 
$R^a R^a$ is proportional to the 
unit matrix: 
\be
R^a R^a = C_R.I 
\ee 
For the adjoint representation we have
\be
C_A \delta_{ab} = f_{acd} f_{bcd}.
\ee
$I_2(R)$ is defined by
\be
\Tr[R^aR^b]= I_2 (R)\delta^{ab}.
\ee 
Thus we have 
\be
C_R d_R = I_2(R) N_A
\ee
where $N_A$ is the number of generators and $d_R$ is the dimensionality 
of the representation $R$. Evidently $I_2 (A) = C_A$.
The fully symmetric 
tensors $d_R^{abcd}$ and
$d_A^{abcd}$ are defined by
\bea
d_R^{abcd}&=&\frac{1}{6}\Tr[R^{(a}R^bR^cR^{d)}],\nn
d_A^{abcd}&=&\frac{1}{6}\Tr[F^{(a}F^bF^cF^{d)}],
\eea
where 
\be
(F^a)^{bc} = if^{bac}
\ee
and
\bea
R^{(a}R^bR^cR^{d)} &=& R^aR^bR^cR^d + R^aR^bR^dR^c 
+R^aR^cR^bR^d\nn  &+& R^aR^cR^dR^b +R^aR^dR^bR^c +R^aR^dR^cR^b, 
\eea
(similarly for $F^{(a}F^bF^cF^{d)}$). 

The additional tensor invariants occurring in
our results for $\beta_s$ and $\gamma_m$ are defined as
\bea
D_2 (A) &=& d_A^{abcd} d_A^{abcd} \nn
D_2 (RA) &=& d_R^{abcd} d_A^{abcd} \nn
D_3 (A) &=& d_A^{abcd} d_A^{cdef} d_A^{abef}\nn
D_3 (RAA) &=& d_R^{abcd} d_A^{cdef} d_A^{abef}.
\eea
In table~\ref{gaminvsa}-\ref{gaminvsc}\ we present results 
for the various tensor invariants
for the groups $SU(N)$, $SO(N)$ and $Sp(N)$, when the fermion 
representation $R$ 
is the fundamental representation. In each case the constant $b$ reflects the 
arbitrariness in the choice of normalisation of the generators 
(see \eqn{eq:sunalg}{} for $SU(N)$). 


\begin{table}[ht]
\begin{center}
\begin{tabular}{|c|c|} \hline
&  \\ 
Group & $SU(N)$ \\ 
&  \\ 
& \\ \hline
 &  \\ 
$C_A$& $bN$ \\ 
& \\ \hline
& \\ 
$C_R$&$b\frac{N^2-1}{2N}$\\
& \\ \hline
& \\ 
$I_2(A)$&$bN$\\
& \\ \hline
& \\ 
$I_2(R)$&$\frac{b}{2}$\\
& \\ \hline
& \\ 
$N_A$ & $N^2-1$ \\ 
& \\ \hline
& \\ 
$D_2 (A)$&$\frac{b^4}{24}(N^2-1)(N^2+36)N^2$
\\
& \\ \hline                   
& \\ 
$D_2 (RA)$&$\frac{b^4}{48}N(N^2-1)(N^2+6)$\\
& \\ \hline
& \\ 
$D_2 (R)$&$\frac{b^4}{96N^2}(N^2-1)(18 - 6N^2 + N^4)$
\\ 
& \\ \hline
& \\ 
$D_3 (A)$&$\frac{b^6}{216}N^2(N^2-1)(324 + 135N^2 + N^4)$\\
& \\ \hline
& \\ 
$D_3 (RAA)$ &$\frac{b^6}{432}N^3(N^2-1)(51 + N^2)$ \\ 
& \\ 
& \\ \hline
\end{tabular}
\caption{\label{gaminvsa} $SU(N)$ Group invariants (here $R$ is the 
fundamental representation).}
\end{center}   
\end{table}

\begin{table}[ht]
\begin{center}
\begin{tabular}{|c|c|} \hline
& \\ 
Group & $SO(N)$ \\ 
& \\ 
& \\ \hline
& \\ 
$C_A$& $b(N-2)$  \\ 
& \\ \hline
& \\ 
$C_R$&$\frac{b}{2}(N-1)$\\
&  \\ \hline
&   \\ 
$I_2(A)$&$b(N-2)$\\
&  \\ \hline
&   \\ 
$I_2(R)$&$b$\\
&  \\ \hline
&   \\ 
$N_A$ & $\frac{1}{2}N(N-1)$ \\ 
& \\ \hline
& \\ 
$D_2 (A)$&$\frac{b^4}{48}N(N-1)(N-2)(-296 + 138N - 15N^2 + N^3)$\\
& \\ \hline                   
& \\ 
$D_2 (RA)$&$\frac{b^4}{48}N(N-1)(N-2)(22 - 7N + N^2)$\\
& \\ \hline
& \\ 
$D_2 (R)$ &$\frac{b^4}{48}N(N-1)(4 - N + N^2)$\\ 
& \\ \hline
& \\ 
$D_3 (A)$&$\frac{b^6}{864}(N-2)(N-1)N(-29440
+23272N- 7018N^2 + 971N^3 - 47N^4 +2N^5)$\\
& \\ \hline
& \\ 
$D_3 (RAA)$ & $\frac{b^6}{864}N(N-2)(N-1)(2048 - 1582N 
+ 387N^2 - 31N^3 + 2N^4)$\\
& \\ \hline
\end{tabular}
\caption{\label{gaminvsb} $SO(N)$ Group invariants (here $R$ is the 
fundamental representation).}
\end{center}   
\end{table}

\begin{table}[ht]
\begin{center}
\begin{tabular}{|c|c|} \hline
& \\ 
Group & $Sp(N)$ \\ 
& \\ 
& \\ \hline
& \\ 
$C_A$& $b(N+2)$  \\ 
& \\ \hline
& \\ 
$C_R$&$\frac{b}{4}(N+1)$\\
&  \\ \hline
&   \\ 
$I_2(A)$&$b(N+2)$\\
&  \\ \hline
&   \\ 
$I_2(R)$&$\frac{b}{2}$\\
&  \\ \hline
&   \\ 
$N_A$ & $\frac{1}{2}N(N+1)$ \\ 
& \\ \hline
& \\ 
$D_2 (A)$&$\frac{b^4}{768}N(N+1)(N+2)(296 + 138N + 15N^2 + N^3)$\\
& \\ \hline                   
& \\ 
$D_2 (RA)$&$\frac{b^4}{768}N(N+1)(N+2)(22 + 7N + N^2)$\\
& \\ \hline
& \\ 
$D_2 (R)$ &$\frac{b^4}{768}N(N+1)(4 + N + N^2)$\\ 
& \\ \hline
& \\ 
$D_3 (A)$&$\frac{b^6}{55296}(N+2)(N+1)N(29440
+23272N+ 7018N^2 + 971N^3 + 47N^4 +2N^5)$\\
& \\ \hline
& \\ 
$D_3 (RAA)$ & $\frac{b^6}{55296}N(N+2)(N+1)(2048 + 1582N 
+ 387N^2 + 31N^3 + 2N^4)$\\
& \\ \hline
\end{tabular}
\caption{\label{gaminvsc} $Sp(N)$ Group invariants (here $R$ is the 
fundamental representation).}
\end{center}   
\end{table}

\section{The groups $SO(N)$ and $Sp(N)$}

In this section we derive explicit expressions for the 
$\beta$-functions for the \epscalar{} quartic interactions for  
the groups $SO(N)$ and $Sp(N)$. 
These may also be derived from Eqs.~(\ref{eq:genbetavs})\ 
using tables~\ref{gaminvsb}, \ref{gaminvsc}.

\subsection{The case ${\cal G} = SO(N)$}

Let us consider $SO(N)$.  
The defining representation of the generators of 
$SO(N)$ is given by the set of matrices 
\be
(M_{[ij]})_{kl}= - i(\delta_{ik}\delta_{jl}-\delta_{il}\delta_{kj}),
\ee
satisfying the algebra 
\be
\left[ M_{[ij]}, M_{[kl]}\right] =  
-i\left(\delta_{jk}M_{[il]}-\delta_{ik}M_{[jl]}
-\delta_{jl}M_{[ik]}+\delta_{il}M_{[jk]}\right)
\ee
or
\be
\left[ M_{[ij]}, M_{[kl]}\right] =  i f_{[ij][kl][mn]} M_{[mn]}
\ee
where the structure constants are given by
\be
f_{[ij][kl][mn]}= \frak{1}{2}\left[
\delta_{ik}\delta_{jm}\delta_{ln}+\ldots (7\hbox{terms})\right]
\ee
such that they are antisymmetric in $ij$, $kl$, and $mn$ exchange.
With this normalisation of the generators it is straightforward to show that 
the adjoint quadratic Casimir $C_A$ is given by 
\be
f_{[ij][kl][mn]} f_{[ij][kl][pq]} = C_A (\delta_{mp}\delta_{nq} - 
\delta_{mq}\delta_{np})
\ee
with 
\be
C_A = 2 (N - 2).
\ee
We will, however, present results for an arbitrary normalisation 
of the generators such that 
\be
C_A = b (N - 2),
\ee
where $b$ is a constant. 
Useful checks on our calculations will be provided by the 
isomorphisms
\be
SO(3) \equiv \frac{SU(2)}{Z_2}
\ee
and
\be
SO(6) \equiv \frac{SU(4)}{Z_2}
\ee
which mean that the Lie algebras of $SO(3)$ and $SU(2)$, and of $SO(6)$ and 
$SU(4)$ are identical.
Note that to compare our result for $SO(3)$ with the 
corresponding result for $SU(2)$ (where with the conventional normalisation 
we have $C_A = 2$) we will need to set $b = 2$, while to compare $SO(6)$ with 
$SU(4)$ we will similarly need to set $b = 1$.

The basis for 4-tensors for $SO(N)$  for $N\geq 4$ has $\gamma = 6$ and 
can be chosen to be\footnote{As will become clear our results will not 
be applicable to the special case $N=8$,  which we will not consider
further}(we adopt a shorthand notation with $[i_1i_2] \to i$ etc.):
\bea
P_1 &=&\delta_{ij}\delta_{kl},\nn
P_2 &=&\delta_{ik}\delta_{jl},\nn
P_3 &=&\delta_{il}\delta_{kj},\nn 
P_4 &=&f_{ijm}f_{klm},\nn
P_5 &=&f_{ikm}f_{jlm},\nn
P_6 &=&\hbox{tr}[F_{i}F_{j}F_{k}F_{l}],\nn
\label{eq:pbasis}
\eea
where $(F_{i})_{mn}=f_{min}$.

Some useful identities for reduction of various 4-tensors to the basis are 
as follows :
\bea
\Tr [F_iF_jF_pF_kF_lF_p] &=& \frac{b^3}{2}(N-4)(-2P_1+2P_2+P_3) 
\nn
&+&\frac{b^2}{4} (N-8)(P_4-2P_5) - \frac{b}{2} (N-2)P_6 
\eea
\bea
\Tr [F_iF_mF_jF_n]
(F_kF_l)_{mn}
&=&\frac{b^3}{2}(N-4)(-2P_1+P_2 + P_3)\nn 
&-&\frac{b^2}{4}(N-8)P_4+\frac{b^2}{2}(N-8)P_5.
\eea
It is interesting that this doesn't involve the basis element $P_6$.
\bea
\Tr [F_iF_jF_mF_n] \Tr [F_kF_lF_mF_n]
&=&\frac{b^4}{4}(N-2)(N-4)(6P_1-P_2 - P_3)\nn
&+&\frac{b^3}{8}(N^3-6N^2+16N-24)P_4\nn 
&-& \frac{b^3}{4}N(N-6)P_5\nn
&+& \frac{b^2}{4}(N^2-6N+20)P_6
\eea
\bea
\Tr [F_iF_mF_jF_nF_kF_mF_lF_n]
&=&\frac{b^4}{4}(N-2)(N-4)(P_1+P_3)\nn
&+&\frac{b^3}{8}(N^2-14N+32)(-2P_4 + P_5)\nn
&-&\frac{b^2}{4}(N-8)P_6
\eea

The results for the one-loop $\beta$-functions in 
a basis as  in Eq.~(\ref{eq:hbarbasis})\ and  with 
$v_i$ defined as for $SU(N)$ are as follows:
\bea
\beta_{v_1}&=&8v_1^2+2(N^2-N+2)v_1v_2
-4b(N-2)v_1v_3+6b^2(N-2)^2v_1v_4\nn 
&+& 12v_2^2 + 8b(N-2)v_2v_3+8b^2(N-2)^2v_2v_4\nn
&+& 20b^4(N-2)(N-4)v_4^2 + 16b^3(N-4)v_3v_4-12b(N-2)v_1\as ,\nn
\beta_{v_2}&=&(N^2-N+8)v_2^2+12v_1v_2-4b(N-2)
v_2v_3+6b^2(N-2)^2v_2v_4\nn
&-&8b^3(N-4)v_3v_4
+6b^4(N-2)(N-4)v_4^2-12b(N-2) v_2\as,\nn
\beta_{v_3}&=&4b(N-2)v_3^2+12v_1v_3 
-4v_2v_3-4b(N-2)v_2v_4\nn 
&+& 2b^2(N^2-6N+20)v_3v_4-4b^3(N-4)v_4^2-12b(N-2) v_3\as ,\nn
\beta_{v_4}&=&\frak12b^2(3N^2-28N+140)v_4^2
+12v_1v_4+20v_2v_4-2v_3^2\nn 
&-& 2b(N-2)v_3v_4 -12b(N-2)v_4\as +6\as^2 .
\label{eq:betavon}
\eea
If we set $N=6$ and $b=1$ in Eq.~(\ref{eq:betavon}), we reproduce
precisely  the results obtained by setting $N=4$ in
Eq.~(\ref{eq:betav}). Similarly,  if we set $N=3$ and $b=2$ in
Eq.~(\ref{eq:betavon}), we reproduce precisely  the results obtained by
setting $N=2$ in Eq.~(\ref{eq:betav}), setting  $v_3 = v_4 = 0$ in both
cases.

\subsubsection{The fermion contribution}

As in the $SU(N)$ case, the fermion loop contribution 
to the scalar anomalous dimension results 
in a contribution of 
$\Delta \beta_{v_i} = 8n_f I_2 (R)   v_i\aep$ to each $\beta$-function 
in Eq.~(\ref{eq:betavon}).
The 1PI fermion box diagram 
makes a contribution to the $\beta$-functions (appropriately normalised) 
of the form as Eq.~(\ref{eq:floop}); for the adjoint representation this 
becomes:
\be
\Hbar_i\Delta\beta_{v_i} =  \aep^2  (-2 b(N-2) \Hbar_3 -4 \Hbar_4),
\ee 
and hence for the complete $\beta$-functions 
including a single two-component fermion in the adjoint representation 
we have from Eq.~(\ref{eq:betavon}):
\bea
\beta_{v_1}&=&8v_1^2+2(N^2-N+2)v_1v_2
-4b(N-2)v_1v_3+6b^2(N-2)^2v_1v_4\nn 
&+& 12v_2^2 + 8b(N-2)v_2v_3+8b^2(N-2)^2v_2v_4+ 16b^3(N-4)v_3v_4\nn
&+& 20b^4(N-2)(N-4)v_4^2 -12b(N-2)v_1\as  +4b(N-2) v_1\aep ,\nn
\beta_{v_2}&=&(N^2-N+8)v_2^2+12v_1v_2-4b(N-2)
v_2v_3+6b^2(N-2)^2v_2v_4\nn
&-&8b^3(N-4)v_3v_4
+6b^4(N-2)(N-4)v_4^2\nn &-& 12b(N-2)v_2\as  + 4b(N-2) v_2\aep ,\nn
\beta_{v_3}&=&4b(N-2)v_3^2+12v_1v_3 
-4v_2v_3-4b(N-2)v_2v_4\nn 
&+& 2b^2(N^2-6N+20)v_3v_4-4b^3(N-4)v_4^2-12b(N-2)v_3\as \nn 
&+& 4b(N-2) v_3\aep  -2b(N-2)\aep^2 ,\nn
\beta_{v_4}&=&\frak12b^2(3N^2-28N+140)v_4^2
+12v_1v_4+20v_2v_4-2v_3^2\nn 
&-& 2b(N-2)v_3v_4 -12b(N-2)v_4\as +6\as^2 \nn  
&+& 4b(N-2)v_4\aep   -4\aep^2, 
\label{eq:betavontot}
\eea
when it is once again easy to extract the \sic\ result
by setting $v_1 = v_2 = v_4 = 0$ and $v_3 = \aep  = \as $.

For a single two-component fermion in the  fundamental representation, (and 
for $N \neq 8$) we find that
\be
\Tr[M_iM_jM_kM_l]=\frac{1}{N-8}\Bigl[
-b^2(P_1 + P_2 +P_3)+b(2P_4 - P_5) + P_6
\Bigr]
\ee  
and hence we find for fermions 
in the fundamental representation a contribution 
to the $\beta$-functions (for $n_f$ flavours) of the form
\be
\Hbar_i\Delta\beta_{v_i} = \frac{2n_f\alpha_e^2}{N-8}
\left[ 8 b^2 (\Hbar_1 + \Hbar_2) -2b(N-10)\Hbar_3-4\Hbar_4\right].
\ee
It is straightforward to incorporate 
these  contributions into Eq.~(\ref{eq:betavon}) in the same manner.

\subsection{The case ${\cal G} = Sp(N)$}

We will here be considering the unitary symplectic group.
The generators of $Sp(N)$ satisfy 
\be
J R^a J = (R^a)^T
\label{eq:spdef}
\ee
where 
\be
J = \left(\begin{matrix}
0&I\\ -I & 0\\
\end{matrix}\right)
\ee
and $I$ is the unit matrix. Evidently $N$ must be even. For the case 
$N=2$ it is easy to check by explicitly constructing $R^a$ to satisfy 
Eq.~(\ref{eq:spdef})\ that $Sp(2)\equiv SU(2)$. 
Another useful check on our calculations will be provided by the 
isomorphism
\be
SO(5) \equiv \frac{Sp(4)}{Z_2}.
\label{eq:so5}
\ee

If we write $N = 2n$, the generators may be written as $L_{\alpha\beta}$,
where an infinitesimal group element ${\cal S}$ may be written 
\be
{\cal S} = 1 + i \sum_{\alpha\beta} a_{\alpha\beta}L_{\alpha\beta}
\ee
where $a_{\alpha\beta} = a^*_{\beta\alpha}$ and 
\be
L_{\alpha\beta} = L_{-\beta-\alpha}, \quad \alpha,\beta = \pm 1, \pm 2, \cdots \pm n
\ee
(Thus the correspondence $Sp(2)\sim SU(2)$ is $L_{11} \sim J_3$, 
$L_{1 -1}, L_{-1, 1} \sim  J_{\pm} = J_1 \pm i J_2$.)

They obey the commutation relations
\be
\left[ L_{\alpha\beta}, L_{\gamma\delta}\right] =  
\left(\delta_{\beta\gamma}\delta_{\beta}L_{\alpha\delta}
-\delta_{\alpha\delta}\delta_{\alpha}L_{\gamma\beta}
+\delta_{\beta\deltab}\delta_{\beta}L_{\alpha\gammab}
-\delta_{\alpha\gammab}\delta_{\alpha}L_{\betab\delta}\right)
\ee
where 
\be
\delta_{\alpha} = - \delta_{-\alpha} = 
\left\{ \begin{array}{ll}
1& \mbox{if $\alpha > 0$},\\
-1& \mbox{if $\alpha <  0$}.
\end{array}\right.
\ee
and $\alphab = -\alpha$ etc.

We find:
\bea
\beta_{v_1}&=& 8v_1^2+2(N^2+N+2)v_1v_2
-2b(N+2)v_1v_3+\frak32b^2(N+2)^2v_1v_4\nn 
&+& 12v_2^2 + 4b(N+2)v_2v_3+2b^2(N+2)^2v_2v_4\nn
&+& \frak54b^4(N+2)(N+4)v_4^2 + 2b^3(N+4)v_3v_4-6b(N+2)v_1\as ,\nn
\beta_{v_2}&=& (N^2+N+8)v_2^2+12v_1v_2-2b(N+2)
v_2v_3+\frak32b^2(N+2)^2v_2v_4\nn
 &-& b^3(N+4)v_3v_4
+\frak38b^4(N+2)(N+4)v_4^2-6b(N+2)v_2\as ,\nn
\beta_{v_3}&=& 2b(N+2)v_3^2+12v_1v_3
-4v_2v_3-2b(N+2)v_2v_4\nn
&+& \frak12b^2(N^2+6N+20)
v_3v_4 - \frak12b^3(N+4)v_4^2-6b(N+2)v_3\as ,\nn
\beta_{v_4}&=& \frak18b^2(3N^2+28N+140)v_4^2
+12v_1v_4+20v_2v_4-2v_3^2\nn 
&-& b(N+2)v_3v_4 - 6b(N+2)v_4\as +6\as^2 .
\label{eq:betasp}
\eea
Using table~\ref{gaminvsc}\ in \eqn{eq:genbetavs}\ leads to 
the same results.

Setting $N=2$, $b = 1$  and $v_3 = v_4 = 0$ in \eqn{eq:betasp}\ above  
we indeed find agreement with the corresponding results for $SU(2)$, 
from Eq.~(\ref{eq:betav}).

\subsubsection{The fermion contribution}

The fermion loop contribution
is similar to the the $SO(N)$ case. 
From the scalar anomalous dimension we get
a contribution of 
$\Delta \beta_{v_i} = 8 n_f I_2 (R)  v_i\aep $ to each $\beta$-function 
in Eq.~(\ref{eq:betasp}).

In the case of the adjoint representation, by similar algebra to 
that leading to Eq.~(\ref{eq:fadj}), we obtain for the 1PI 
fermion box diagram a contribution: 
\be
\Hbar_i\Delta\beta_{v_i} =  \aep^2  (-b(N+2) \Hbar_3 -4 \Hbar_4),
\ee 
and for the complete $\beta$-functions from Eq.~(\ref{eq:betasp}):
\bea
\beta_{v_1}&=& 8v_1^2+2(N^2+N+2)v_1v_2
-2b(N+2)v_1v_3+\frak32b^2(N+2)^2v_1v_4\nn 
&+& 12v_2^2 + 4b(N+2)v_2v_3+2b^2(N+2)^2v_2v_4\nn
&+& \frak54b^4(N+2)(N+4)v_4^2 + 2b^3(N+4)v_3v_4-6b(N+2)v_1\as \nn
&+& 2b(N+2)v_1\aep ,\nn
\beta_{v_2}&=& (N^2+N+8)v_2^2+12v_1v_2-2b(N+2)
v_2v_3+\frak32b^2(N+2)^2v_2v_4\nn
 &-& b^3(N+4)v_3v_4
+\frak38b^4(N+2)(N+4)v_4^2-6b(N+2)v_2\as \nn &+&2b(N+2)v_2\aep ,\nn
\beta_{v_3}&=& 2b(N+2)v_3^2+12v_1v_3
-4v_2v_3-2b(N+2)v_2v_4\nn
&+& \frak12b^2(N^2+6N+20)
v_3v_4 - \frak12b^3(N+4)v_4^2-6b(N+2)v_3\as ,\nn
&+& 2b(N+2)v_3\aep  - b(N+2)\aep^2 ,\nn 
\beta_{v_4}&=& \frak18b^2(3N^2+28N+140)v_4^2
+12v_1v_4+20v_2v_4-2v_3^2\nn 
&-& b(N+2)v_3v_4 - 6b(N+2)v_4\as +6\as^2 \nn &+&2b(N+2) v_4\aep  - 4\aep^2 ,
\label{eq:betasptot}
\eea
when it is once again easy to extract the \sic\  result
by setting $v_1 = v_2 = v_4 = 0$ and $v_3 = \aep  = \as $.
Setting $N=4$ in Eq.~(\ref{eq:betasptot})\ we reproduce 
precisely the results of setting $N=5$ in Eq.~(\ref{eq:betavon}), 
in accordance with Eq.~(\ref{eq:so5}); 
a good check of our calculation. Also, setting $N=2$ 
and $v_3 = v_4 = 0$ in 
Eq.~(\ref{eq:betasptot})\ we reproduce the results of setting 
$N=2$ and $v_3 = v_4 = 0$ in Eq.~({\ref{eq:betav}).

For a single two-component fermion in the  fundamental representation,
we find (again reverting to a shorthand single index notation) 
\be
\Tr[L_{\alpha}L_{\beta}L_{\gamma}L_{\delta}]=\frac{1}{N+8}\Bigl[
-\frac{b^2}{4}(P_1 + P_2 +P_3)-\frac{b}{2}(2P_4 - P_5) + P_6
\Bigr], 
\label{eq:tracesp}
\ee 
(where the $P$-basis is defined in the same way as for $SO(N)$ 
in Eq.~(\ref{eq:pbasis}))
and hence a contribution 
to the $\beta$-functions (for $n_f$ flavours) of the form
\be
\Hbar_i\Delta\beta_{v_i} = \frac{2n_f\alpha_e^2}{N+8}
\left[ 2 b^2 (\Hbar_1 + \Hbar_2) -b(N+10)\Hbar_3-4\Hbar_4\right].
\label{eq:spferm}
\ee
It is straightforward to incorporate 
these  contributions into Eq.~(\ref{eq:betasp}) in the same manner.

We can also check this result \footnote{Of course we cannot use
Eq.~(\ref{eq:so5}) as a check here  because the fundamental
representation is different for the two groups.}  using the identity 
$Sp(2)\equiv SU(2)$; setting $N=2$ and $b=1$ in Eq.~(\ref{eq:spferm}) 
and using Eq.~(\ref{eq:hbarsu2}) we find agreement with the result of 
setting $N=2$ and using  Eq.~(\ref{eq:hbarsu2}) in
Eq.~(\ref{eq:suferms}).  

Remarkably, the all $\beta$-functions for the $Sp(N)$ case
(including $\beta_s$, $\beta_e$) together with $\gamma_m$  
can be derived from the corresponding
$SO(N)$ versions by a series of simple substitutions:
\bea
b &\to& \frak{1}{2}b\nn
N  &\to& -N\nn
\alpha_s&\to& -\alpha_s\nn
\alpha_e&\to& -\alpha_e\nn
v_3 &\to& -v_3\nn
n_f &\to& -n_f.
\label{eq:sptrans}
\eea

\vspace*{1em}

\noindent
{\large\bf Acknowledgements}\\ 
One of us (DRTJ) thanks KITP (Santa Barbara) for financial support and 
hospitality while part of this work was done. PK and LM thank Robert
Harlander and Matthias Steinhauser for valuable discussions and explanations. 
This work was  supported by the DFG through SFB/TR~9  and
by  the National Science Foundation   under Grant No. PHY05-51164.

\end{document}